\documentclass[12pt]{spieman}  %>>> use for US letter paper
\usepackage{amsmath,amsfonts,amssymb}
\usepackage{graphicx}
\graphicspath{{figures/}}
\usepackage[colorlinks=true, allcolors=blue]{hyperref}
\usepackage{comment}
\usepackage{setspace}
\usepackage{tocloft}
\usepackage{lineno}
\usepackage{xcolor}
\usepackage[normalem]{ulem}
\usepackage{amsmath}

\title{Laboratory Demonstration of Real-Time Focal Plane Wavefront Control of Residual Atmospheric Speckles}

\author[a,b,*]{Benjamin L. Gerard}
\author[b,a]{Daren Dillon}
\author[c,b]{Sylvain Cetre}
\author[b,a]{Rebecca Jensen-Clem}

\affil[a]{University of California Observatories, CA, USA}
\affil[b]{University of California Santa Cruz, CA, USA}
\affil[c]{W.M. Keck Observatory, HI, USA}

\cftpagenumbersoff{figure}
\cftpagenumbersoff{table} 
\begin{document} 
\maketitle

% Option to view page numbers
\pagestyle{plain} % change to \pagestyle{plain} for page numbers   

\begin{abstract}
Current and future high contrast imaging instruments aim to detect exoplanets at closer orbital separations, lower masses, and/or older ages than their predecessors. However, continually evolving speckles in the coronagraphic science image limit contrasts of state-of-the-art ground-based exoplanet imaging instruments. For ground-based adaptive optics (AO) instruments it remains challenging for most speckle suppression techniques to attenuate both the dynamic atmospheric as well as quasi-static instrumental speckles on-sky. We have proposed a focal plane wavefront sensing and control algorithm to address this challenge, called the Fast Atmospheric Self-coherent camera (SCC) Technique (FAST), which in theory enables the SCC to operate down to millisecond timescales even when only a few photons are detected per speckle. Here we present the first experimental results of FAST on the Santa Cruz Extreme AO Laboratory (SEAL) testbed. In particular, we illustrate the benefit of ``second stage'' AO-based focal plane wavefront control, demonstrating up to 5x contrast improvement with FAST closed-loop compensation of evolving residual atmospheric turbulence---both for low and high order spatial modes---down to 20 millisecond-timescales.
\end{abstract}

% Include a list of keywords after the abstract 
\keywords{Wavefront Sensing, Wavefront Control, Coronagraphy, Adaptive Optics}

% Include email contact information for corresponding author
{\noindent \footnotesize\textbf{*}Benjamin L. Gerard,  \linkable{blgerard@ucsc.edu} }

\begin{spacing}{2}   % use double spacing for rest of manuscript

\section{INTRODUCTION}
\label{sec:intro}  % 
Current state-of-the-art instrumentation enables ground-based exoplanet imaging instruments to be sensitive to detecting self-luminous giant exoplanets down to a few Jupiter masses at separations beyond around 10 au within star systems that are within around 100 pc and are younger than a few hundred Myrs\cite{gpies}. Despite exciting new detections enabled by these instruments (e.g., Ref. \citenum{51eri}), large surveys have both (a) shown that the giant exoplanets these facilities are sensitive to detecting are rare---as low as 1\% of stars host such planets\cite{idps}---and as a result (b) illustrated that future instruments will need to improve detection and characterization sensitivity to lower mass, closer-in, and older systems. The main technological factor limiting such improved sensitivity for near-infrared imaging is un-corrected/subtracted speckle noise.\cite{fast_phd} Speckle noise is a result of leftover diffracted starlight in the coronagraphic science image that prevents detecting exoplanets below a flux ratio (i.e., planet flux normalized to the host star flux), of about $10^{-6}$ at radial separations at 10$\lambda/D$ from the star\cite{51eri}. It is therefore crucial to improve on these speckle subtraction and correction techniques to enable both current and the next generation of exoplanet imagers to detect and characterize new exoplanetary systems, shedding light on the processes of planet formation, evolution, and ultimately the prevalence of life beyond the Solar System\cite{psi}.

Focal plane wavefront sensing and control\cite{cdi_rev} for ground-based 8m-class telescopes---using either an active feedback loop between the coronagraphic science image and adaptive optics (AO) system's deformable mirror (DM) and/or via post-processing methods (which in this paper we will refer to as coherent differential imaging, or CDI)---is one such speckle correction technology that could enable the aforementioned sensitivity/contrast gains needed for fainter exoplanet detection and characterization. However, minimal on-sky gains with this approach have been demonstrated thus far\cite{scexao_nulling, scexao_ldfc}, largely due to speckle evolution occurring on timescales faster than measurement and correction algorithms and/or system hardware can enable\cite{fast_phd}. However, a number of projects are now being pursued to further develop the promise of this technology on-sky, including ongoing efforts at Keck/NIRC2\cite{keck_speckle_nulling}, Subaru/SCExAO\cite{scexao_ldfc,scexao_mec}, Magellan/MagAO-X\cite{magao_cdi}, ESO/SPHERE\cite{sphere_efc}, and Gemini/GPI\cite{cal2}. In this paper we will discuss related laboratory developments of one such technology, based on the self-coherent camera (SCC)\cite{scc_orig}, called the Fast Atmospheric SCC Technique (FAST) \cite{fast18,fast_phd}, using the Santa Cruz Extreme AO Laboratory (SEAL)\cite{seal}. 
This FAST concept was initially developed at the high contrast testbed at the National Research Council of Canada Herzberg Astronomy and Astrophysics research center, which has published a laboratory optical design and preliminary FAST testing results in Ref. \citenum{ne}. 
A conference proceeding publication of preliminary SEAL testing and development that led to this manuscript was published initially in Ref. \citenum{fast_spie21}. 

We summarize the FAST concept in \S\ref{sec: fast_sum} (with additional background given in Appendix \ref{sec: scc_sum}), provide an overview of our FAST SEAL setup in \S\ref{sec: setup_and_alignment}, and describe various FAST calibration wavefront reconstruction procedures in \S\ref{sec: calib}. We present results and analysis of closed-loop FAST correction in \S\ref{sec: closed_loop}, including for static dark hole generation (\S\ref{sec: static}), quasi-static on-air aberrations (\S\ref{sec: quasi_stat}), and AO residual turbulence (\S\ref{sec: dynamic}), clearly demonstrating the gain that real-time focal plane wavefront control enables. We then provide further discussion in \S\ref{sec: improvement} and conclude in \S\ref{sec: conclusion}.
\section{FAST SUMMARY}
\label{sec: fast_sum}
\begin{figure}[!h]
\centering
\includegraphics[width=0.8\textwidth]{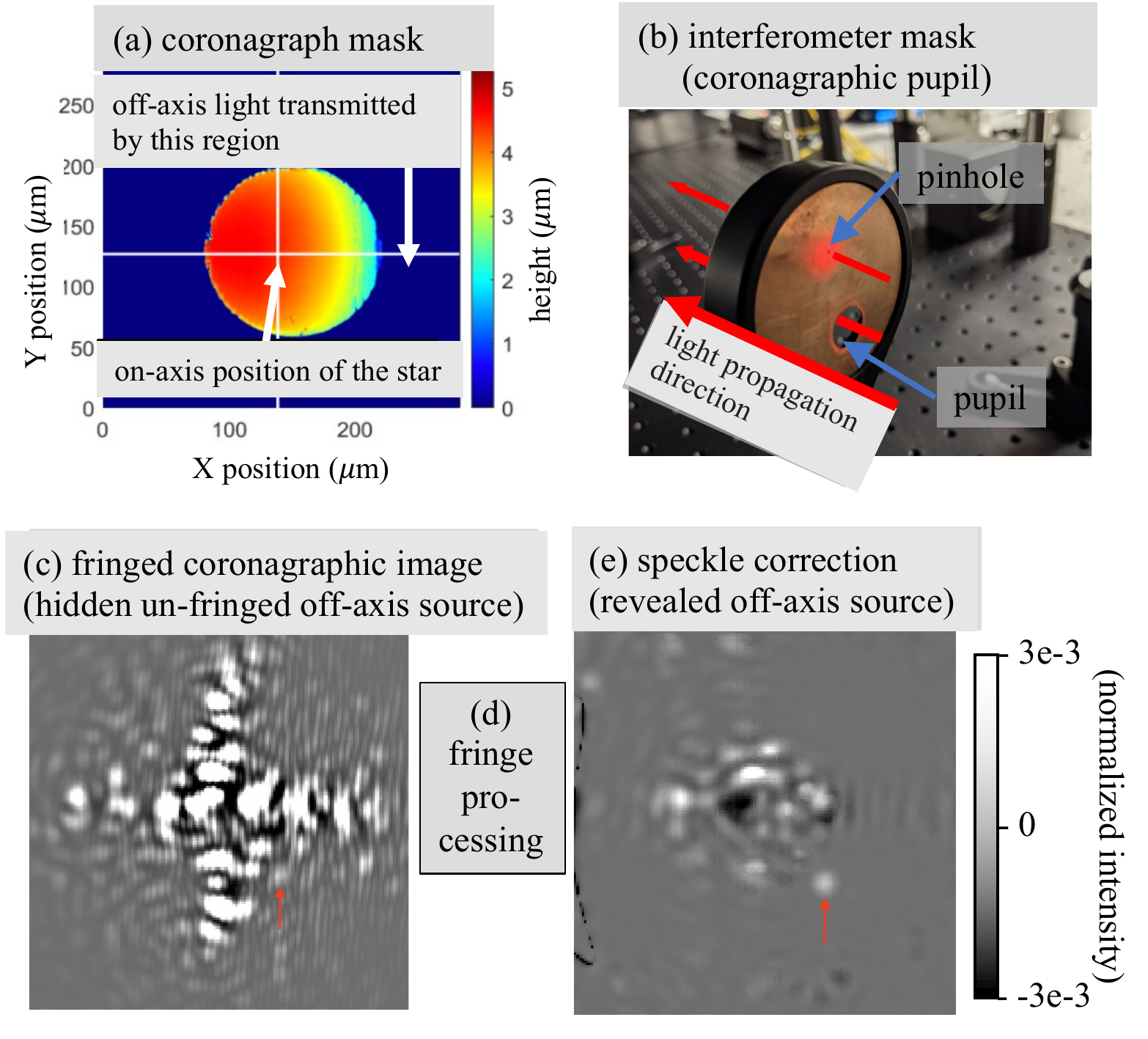}
%\vspace*{-10pt}
\caption[FAST Summary]{Overview of FAST (from Ref. \citenum{fast_spie21}) using images from our Santa cruz Extreme AO Laboratory (SEAL) high contrast testbed\cite{seal} (these FAST/SEAL laboratory results will be presented throughout this paper and are shown here to first illustrate the FAST concept). A setup-specific coronagraph mask (a) is designed to both (i) attenuate starlight while transmitting exoplanet light, and (ii) enhance the stellar fringe visibility using an interferometer mask in the downstream coronagraphic pupil (b). The FPM shape, as shown in panel a, includes a Gaussian and tilt component, designed to redistribute starlight into the pinhole aperture in panel b that is within 3 $\lambda/D$ of the optical axis in panel a, and a flat/piston component that sends star and exoplanet light beyond 3 $\lambda/D$ in panel a into the pupil aperture in panel b. This design enables the detection of an interference pattern on top of the leftover stellar speckles in the coronagraphic image (c), enabling coronagraphic science imaging and AO wavefront sensing. The fringed stellar speckles are then processed (d) and further attenuated using both a deformable mirror and via post-processing  (e), revealing an un-fringed off-axis source. See \S\ref{sec: calib} and \S\ref{sec: closed_loop} for more details on panels d and e.\vspace{10pt}}
\label{fig: fast_sum}
\end{figure}

FAST, illustrated and described in Fig. \ref{fig: fast_sum}, relies on a technique in which coherent starlight is interfered with itself to enable a measurement and correction of the complex speckle electric field (i.e., phase and amplitude) in the coronagraphic image. 
See Appendix \ref{sec: scc_sum} for a detailed description of of the self-coherent camera (SCC)\cite{scc_orig}, upon which FAST is based. Speckle attenuation/subtraction with this approach can be accomplished using a DM (i.e., focal plane wavefront control)\cite{fast_spie18}, and/or by post-processing (i.e., CDI)\cite{fast18, fast_spie21}. In this paper, we will only focus on FAST DM control.
As shown in Fig. \ref{fig: fast_sum}, FAST produces stellar interference fringes by way of two specialized coronagraphic masks: a setup-specific focal plane mask (Figure \ref{fig: fast_sum}a, illustrating a tip+Gaussian component within 3 $\lambda/D$ and a piston component elsewhere) modulates starlight but not exoplanet light such that the downstream pupil plane mask (Figure \ref{fig: fast_sum}b) transmits remaining starlight (i.e., including residual aberrations from the atmosphere, telescope and instrument optics, and coronagraph mask diffraction effects) and exoplanet light through a traditional Lyot stop, while much of the starlight (but not exoplanet light) passes through a separate pinhole in the mask. In the final focal plane (the coronagraphic image), the residual stellar speckles interfere with the starlight that passed through the pinhole to produce fringes (Figure \ref{fig: fast_sum}c), in principle even when only a few photo-electrons ($\gtrsim$10) are detected per pixel\cite{fast18}, while any off-axis sources remain un-fringed). FAST simulations have shown that Fourier-based processing of the fringes in this final image (see Appendix \ref{sec: scc_sum} for details of this process) enable estimating and attenuating (via DM control and/or CDI-based post-processing) remaining stellar speckles from a single, millisecond-exposure image, with no need to create additional phase diversity by defocusing and/or probing individual speckles with the DM (Figure \ref{fig: fast_sum}e). 

FAST's key innovation compared with previous focal plane wavefront control techniques is the coronagraphic focal plane mask (FPM; Fig. \ref{fig: fast_sum}a), which provides a high enough fringe visibility (i.e., fringed vs. un-fringed components) in the final image such that fringes can in theory be detected on millisecond timescales, even when only a few photons are detected per pixel. The classical SCC also generates fringes in the coronagraphic image to enable focal plane wavefront control and CDI \cite{scc_orig}, but with fringe visibilities about 10$^6$ times lower than Fig. \ref{fig: fast_sum}c (e.g., see Fig. \ref{fig: scc}b in Appendix \ref{sec: scc_im}), removing any possibility of AO-based wavefront control on millisecond timescales. Also due to the nature of the FAST FPM, which sends the core of the PSF (i.e., low order modes of the entrance pupil wavefront) into the off-axis Lyot stop pinhole, in turn transmitted through to the SCC image, on-axis FAST images are highly sensitive to detecting low order modes of the entrance pupil wavefront\cite{fast_spie20}, while typically other coronagraphs usually used with the SCC are comparatively not as sensitive to such modes (see \S\ref{sec: lowfs}).
\section{SEAL LABORATORY SETUP AND ALIGNMENT}
\label{sec: setup_and_alignment}
\subsection{Setup}
\label{sec: setup}
In this section we will describe our refractive setup at UCSC's Laboratory for AO facilities developed to test FAST, which is the same setup from which subsequent results in this paper are obtained. Although we will refer to this setup as the Santa cruz Extreme AO Laboratory (SEAL), SEAL will ultimately be a mostly reflective setup designed for multi-purpose AO-based wavefront sensing and control techniques, beyond the scope of just testing FAST; the higher-level setup of the full SEAL testbed, including the optical design and both current and ongoing projects (including FAST), are presented and described in Ref. \citenum{seal}. Note that the main facilities used here are the same as previously described in Ref. \citenum{lao_granite}, including a highly-stabilized granite testbed and custom-made enclosure.

\begin{figure}[h]
\centering
\includegraphics[width=1.0\textwidth]{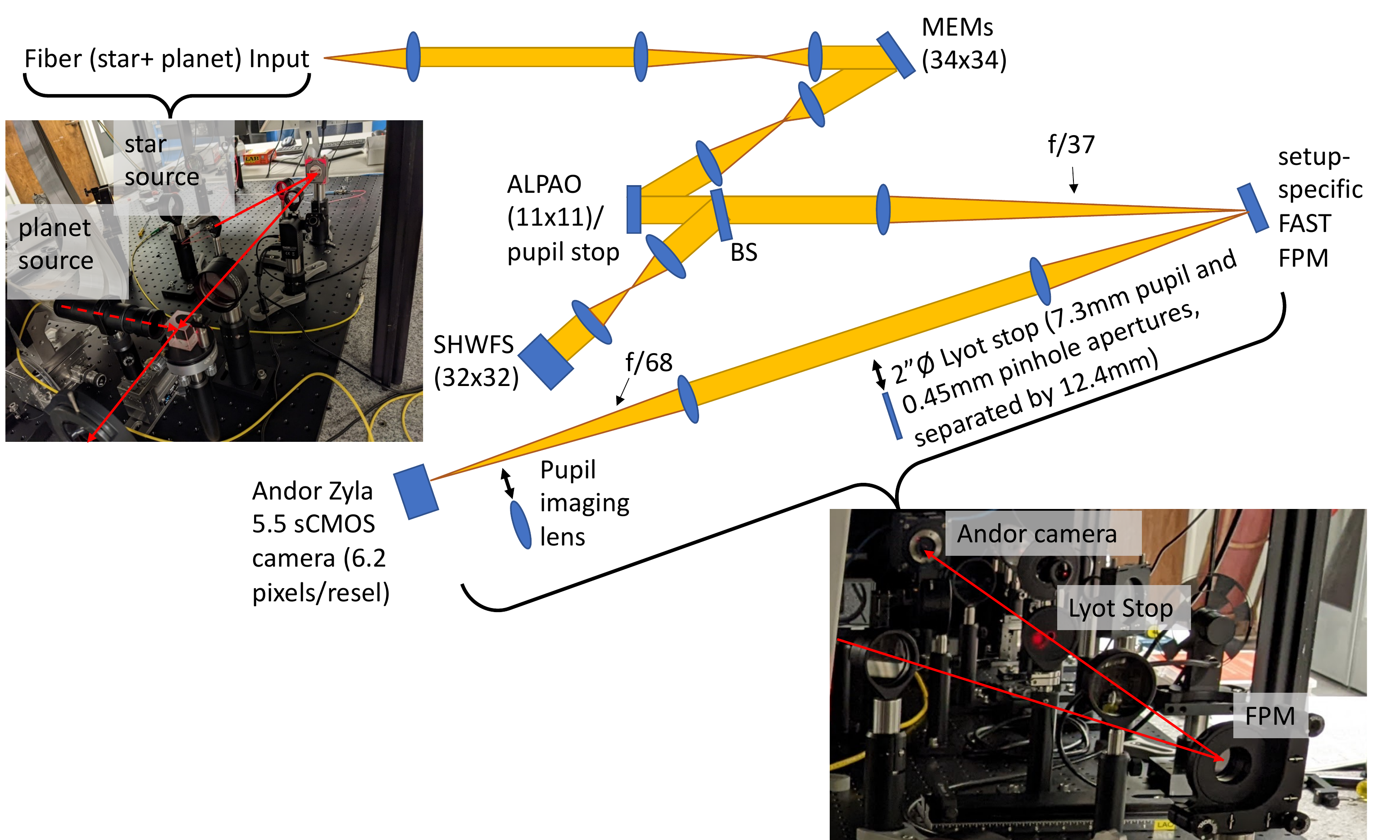}
\caption[SEAL testbed setup]{An outline of our FAST SEAL setup, showing a conceptual diagram of the full setup and an inset of testbed images in two areas. In the inset, red arrows indicate light propagation (with different line styles combined/split from/into different paths). Double sided black arrows indicate a deployable mechanism. Starting in the upper left, a planet and star light source, both aligned to the same focal plane, are collimated and then combined, with the planet source positioned slightly off-axis a few diffraction bandwidths from the star. Re-imaging optics place the pupil first on the 29x29 (across the beam) MEMS DM, and then on the 11x11 ALPAO DM, the latter defining the system pupil stop. A plate beam splitter (labeled as ``BS'') reflects light to a Shack Hartmann WFS (SHWFS) with 32 subapertures across the beam diameter and transmits light to the FAST arm. The FAST beam is then focused on our setup-specific FPM, then re-collimated to a downstream coronagraphic pupil, where a deployable Lyot stop can modulate the beam (the central and off-axis coronagraphic pupil are visible on the Lyot stop in the lower left inset). Finally, downstream imaging optics enable coronagraphic focal plane (i.e., FAST) or pupil plane imaging on an Andor camera.}
\label{fig: lab_setup}
\end{figure}

Fig. \ref{fig: lab_setup} shows our SEAL testbed setup and outlines the main hardware components relevant to FAST testing.
The visible light star and planet light sources are commercial Thorlabs lasers KLS635 and HNLS008L, respectively, centered at 633 nm. The adjustable star light source is set to 0.15 mW unless otherwise noted, while the planet light source is fixed at 0.8 mW. We place a neutral density (ND) filter with an optical density (OD) of 2 on the planet source. The beamcube combining the two light sources further attenuates the planet by a factor of 9, theoretically setting the planet at a flux ratio of 6$\times10^{-3}$ (see \S\ref{sec: calib} for a measurement of the star-to-planet flux ratio).
\footnote{Although for most systems this flux ratio is unlikely to yield planet-mass sensitivity (i.e., more realistically in the brown dwarf mass sensitivity range), for the purposes of this paper (i.e., demonstrating real-time correction of AO residuals, not demonstrating long-exposure high-contrast capabilities) we will subsequently refer to this off-axis source as the planet light source.}
Additional higher OD ND filters (and/or increasing the star light source power) can further decrease this flux ratio for future high contrast experiments, but for the purposes of our main focus in this paper (i.e., real-time FAST correction of AO residuals) the above-described planet flux ratio is sufficient. Moving downstream, a set of relay lenses then images the system pupil (note that the system pupil stop is defined further downstream; see below) onto our 34x34 actuator micro-electrical-mechanical system deformable mirror (MEMS; 952 actuators forming a circular pupil), illuminating 29 actuators across the system beam diameter (but see the Lyot stop sizing implications of the MEMS footprint two paragraphs below). Re-imaging optics then relays a pupil plane to a 11x11 actuator ALPAO DM (97 actuators forming a circular pupil), also serving as the primary system pupil stop (i.e., a circular un-obscured pupil geometry; back-propagation determines the relevant illuminated pupil on the MEMS). A plate beam splitter then reflects 50\% of the beam through a pair of relay lenses to a commercial Thorlabs Shack Hartmann wavefront sensor (SHWFS; model WFS-20), sampling 32 lenslets across the beam diameter. 
The transmitted light is sent though a 500mm focal length lens to generate a f/37 beam on the FPM, which is designed and fabricated specifically for that f-number and the 633 nm light source wavelength. 

Fig. \ref{fig: fast_sum}a and \ref{fig: lab_setup} already show the fully reflective setup-specific FAST Tip/tilt Gaussian (TG) FPM fabricated and characterized via Zygo interferometry measurements by University of Alberta's nanoFAB laboratory, made from a Nanoscribe 3D printing machine and subsequently aluminum coated. This mask design only deviates from a flat surface over the central region (a 3 $\lambda/D$ radius from the optical axis) that redistributes the core starlight, and with this 3D printing approach the full tilt can be made without phase wrapping (i.e., avoiding additional related chromatic effects). However, the design is also essentially a Lyot coronagraph, with no component optimized for diffracted starlight suppression, and so it is not ideal for reaching high contrasts. However, such a mask is still sufficient for the focus of this manuscript (i.e., ground-based observations where coronagraphic images are dominated by un-pinned speckle noise). Further characterization of this mask is presented in Ref. \citenum{fast_spie21}, which shows that it is fabricated at sufficient quality to enable high fringe visibility SCC images to enable real-time wavefront control, which is the main focus of this manuscript.

A f=200mm, 2"{\O} lens then collimates the post-FPM beam, oversizing the clear aperture relative to the non-coronagraphic pupil footprint by a factor of 6.2, sufficient for the ``classical'' SCC\cite{galicher_scc}. A Lyot stop then transmits light through the central 7.3mm coronagraphic pupil (90\% undersized) and off-axis 0.45 mm pinhole (which is the maximum pinhole diameter required to enable the first pinhole PSF Airy minimum to lie outside the DM control region). The 90\% undersized Lyot stop pupil means that the 29 actuators sampled across the non-coronagraphic pupil becomes only 26 actuators sampled across the coronagraphic pupil. Avoiding edge effects, we found that using 24 actuators across the MEMS pupil diameter enabled sufficient performance in the FAST coronagraphic image. Although the FAST FPM prescription is designed for a theoretical pinhole-pupil separation (center to center) of 1.6 pupil diameters, the separation is instead empirically determined to be 1.524 pupil diameters (see \S\ref{sec: fpm_alignment}). A f=500mm 2"{\O} lens lastly focuses the post-Lyot stop beam onto our Andor Zyla 5.5 sCMOS camera, enabling a theoretical plate scale of 6 pixels/resel (see \S\ref{sec: appendix_calibration} for a corresponding measurement), sufficiently over-sampled for the SCC fringes\cite{galicher_scc}. A f=50mm lens is placed 50mm from the Andor detector on a flip mount to reimage the coronagraphic pupil on the same detector when deployed to be used for calibration purposes. The Lyot stop is also on a flip mount to enable coronagraphic pupil imaging both with and without the Lyot stop.

Our SEAL software infrastructure is based on the Keck pyramid wavefront sensor real-time control (RTC) architecture, described in Ref. \citenum{krtc}. Ref. \citenum{seal} describes more specifically the hardware and infrastructure for SEAL software control and development, which includes a GPU-based clone of the Keck RTC in Linux (which we will refer to as the SEAL RTC) and a Windows machine for some off-the-shelf components without available Linux drivers. Low-level drivers for the MEMS and Andor camera (i.e., to get and send DM commands and to get images/sub-arrays) are written on the SEAL RTC in C and interfaced with Python to enable high-level FAST software development. Unless otherwise mentioned, for the Andor camera we use a 320x320 sub-array updating at 100 Hz (although in future work we will plan to demonstrate the higher-speed FAST closed-loop control: Andor benchmarks the Zyla 5.5 with less than 1 e- read noise at frame rates up to 1 kHz; Andor, private communication). Running in serial mode in Python (i.e., not parallelized), a 20 ms pause between grabbing Andor frames and applying DM commands is needed to enable a given image to be used to compute the corresponding DM commands. However, benchmarking tests for other projects on the SEAL RTC have already demonstrated the ability to reach $\sim$5 ms latency in Python by multi-threading and synchronizing imaging-grabbing and DM-command-sending threads.\cite{sengupta21}
%; although we are working to decrease this latency by multi-threading and synchronizing imaging-grabbing and DM-command-sending threads, we will ultimately implement a high-speed high-level FAST RTC in C, analogous to the above-described Keck pyramid wavefront sensor RTC and enabling computational latencies closer to 1 ms (consistent with FAST CPU-based latencies computed in Ref. \citenum{fast_spie18}). 
Our windows machine is used to operate the adjustable star light source, deployable pupil imaging lens, %optical chopper wheel, 
and SHWFS. 
%The optical chopper wheel is electronically synchronized to the Andor camera readout frame rate, using the Andor as the ``leader'' and the chopper as the ``follower,'' manually adjusting the chopper phase to ensure the Lyot stop pinhole in every other frame is blocked and then unblocked. 
For the SHWFS, although Thorlabs does not provide Linux-based drivers for the commercial WFS-20, we use the Windows serial interface in Python and an ethernet connection to the SEAL RTC to use SHWFS frames in our SEAL RTC Python interface for high-level AO software development (as with the Andor and MEMS above); current frames are limited to updating at 50 Hz (i.e., limited by the above-mentioned 20 ms pause) to interface with Python. However, separate benchmarking with this SHWFS setup shows that we can readout slopes with 23 subapertures across the beam (i.e., slightly undersampled relative to the 29 DM actuators across the beam) at 460 Hz for future ``first stage AO + FAST'' tests at higher speeds (see \S\ref{sec: conclusion}).
\subsection{Alignment}
\label{sec: fpm_alignment}
Off-the-shelf lenses are aligned with standard techniques, using a shear plate to collimate the beam and a knife-edge test to calibrate focal lengths, but will not be discussed in detail further in this manuscript. The FAST FPM is first aligned manually using a series of stage adjustments and then algorithmically using the ALPAO DM. Further details of this FPM alignment process are given in Appendix \ref{sec: appendix_fpm_alignment}. Appendix \ref{sec: appendix_lyot_design} additionally describes the steps taken after FPM alignment to empirically define the Lyot stop prescription from recorded coronagraphic pupil images.
\section{CALIBRATION PROCEDURE}
\label{sec: calib}
In this section we describe the process and measurements for high and low order reconstruction (\S\ref{sec: howfs_recon} and \ref{sec: lowfs}, respectively). This analysis relies on a number of initial calibration steps, including ALPAO DM flattening, DM-to-wavefront conversion, and more, which are described in detail in Appendix \ref{sec: appendix_calibration}.
\subsection{High Order Reconstruction}
\label{sec: howfs_recon}
Our high order wavefront calibration and reconstruction procedure largely follows the standard methods developed for calibrating the SCC
as outlined in Appendix \ref{sec: scc_dm}. In addition to the overview given in \S\ref{sec: scc_dm}, we briefly summarize these steps here, specifically noting additional setup-specific subtleties we have implemented. Low order WFS (LOWFS) calibration and linearity will be discussed next in \S\ref{sec: lowfs}.

To generate the high order interaction matrix (IM), sine and cosine waves are applied to the MEMS for every spatial frequency greater than 5 cycles per pupil (c/p) within the DM control region at intervals of 1 c/p. Applying lower order Fourier modes (i.e., those that are close to or within the edge of the FAST FPM at 3 $\lambda/D$) in the same IM as these mid to high order modes would increase non-linearities by modulating the pinhole PSF throughput for only those modes. To minimize aliasing effects we limit the highest MEMS spatial frequency along the x and y direction to 12 c/p, below the theoretical 13 c/p maximum\cite{fast_spie18,fast_spie20}. For each DM Fourier mode, we compute $I_-$ (i.e., the Fourier-filtered complex-valued fringe component of the SCC image, as described in \S\ref{sec: scc_im}, Eq. \ref{eq: I_minus}, and Fig. \ref{fig: scc_wfsing}) pixel values (real and imaginary) within a user-defined pre-computed ``dark hole'' (DH) region (i.e., which sets the 12 c/p maximum spatial frequency limit along the x direction described above) for a differential image (i.e., as in Fig. \ref{fig: scc_IM}). As in Ref. \citenum{fast_spie20}, we implement an additional binary mask before storing the computed $I_-$ for a given mode in the IM, which isolates the given $I_-$-plane sine spot within a 1.3 $\lambda/D$ radius of it's central location and sets all other pixel values within the DM control region equal to zero, minimizing cross-talk between modes. Using the same notation as \S\ref{sec: scc_dm}, looping through all Fourier modes, these $I_-$ pixel values are then stored in a reference vector, $\vec{A}$, with dimensions $n\times m=$ 456$\times$27936. The 456$\times$27936 command matrix (CM), $\vec{C}$, and reconstructed high order DM commands, $\vec{D_n}$, are then generated via equations \ref{eq: cm} and \ref{eq: dmc}, respectively.

We compute the pseudo inverse via singular value decomposition (SVD), choosing the lowest SVD cutoff value (i.e., the lowest number of IM Eigenmodes conserved in the CM) which sufficiently reconstructs individual Fourier modes applied on the DM without cross-talk from other modes. Note that although with a more conventional zonal-based pupil plane WFS where the SVD cutoff is used to minimize the propagation of waffle modes onto the DM\cite{hardy}, this waffle mode propagation is less relevant for a focal plane WFS. A binary DH mask is spatially filtering modes that the DM controls, inherently removing waffle modes at the DM Nyquist frequency\cite{fast_spie20}.
\subsection{LOWFS Reconstruction and Linearity}
\label{sec: lowfs}
Using FAST, the SCC image is also a LOWFS, as demonstrated in simulations in Ref. \citenum{fast_spie20}. Here we demonstrate laboratory validation of this concept with SEAL. FAST LOWFS capabilities are enabled by the setup-specific FAST FPM, which sends the PSF core (i.e., low order pupil plane wavefront aberrations) through the Lyot stop pinhole aperture, forming fringes in the SCC image that track the amplitude and phase of low order modes (including for even modes such as focus, whose sign ambiguity is resolved from the fringes).

\begin{figure}[!h]
    \centering
    \includegraphics[width=0.7\textwidth]{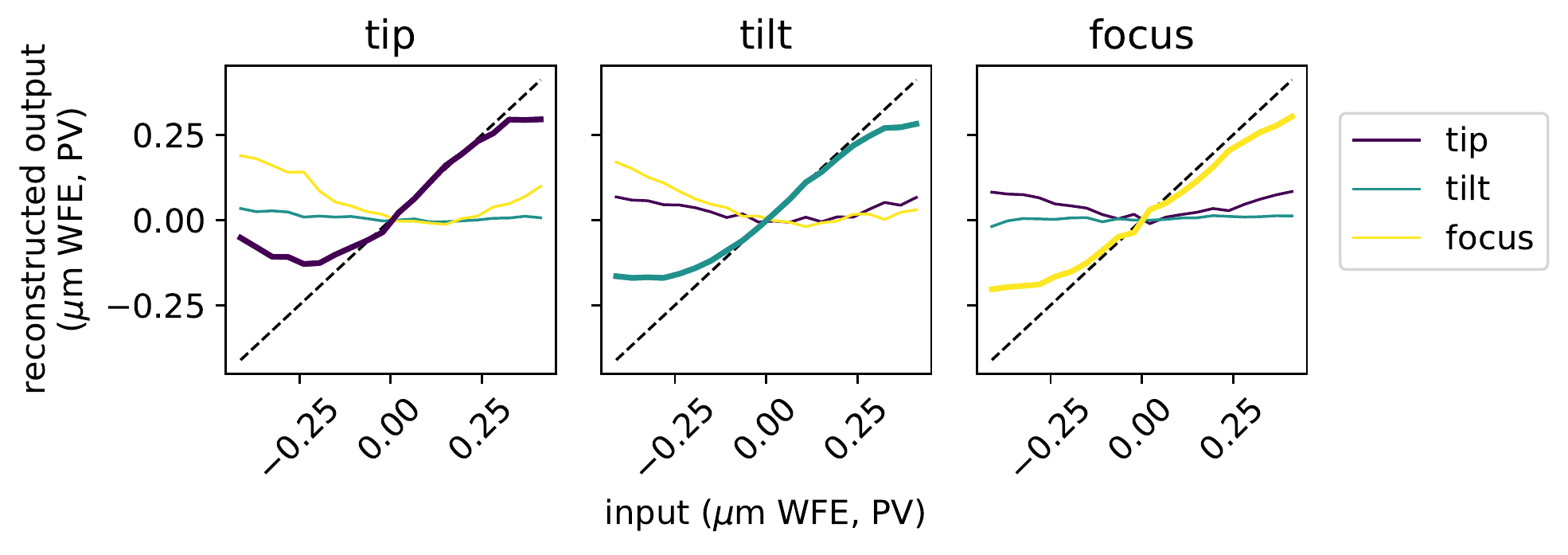}
    \caption[LOWFS linearity for tip/tilt and focus]{LOWFS linearity for tip/tilt and focus.}
    \label{fig: ttf_linearity}
\end{figure}

\begin{figure}[!h]
    \centering
    \includegraphics[width=0.7\textwidth]{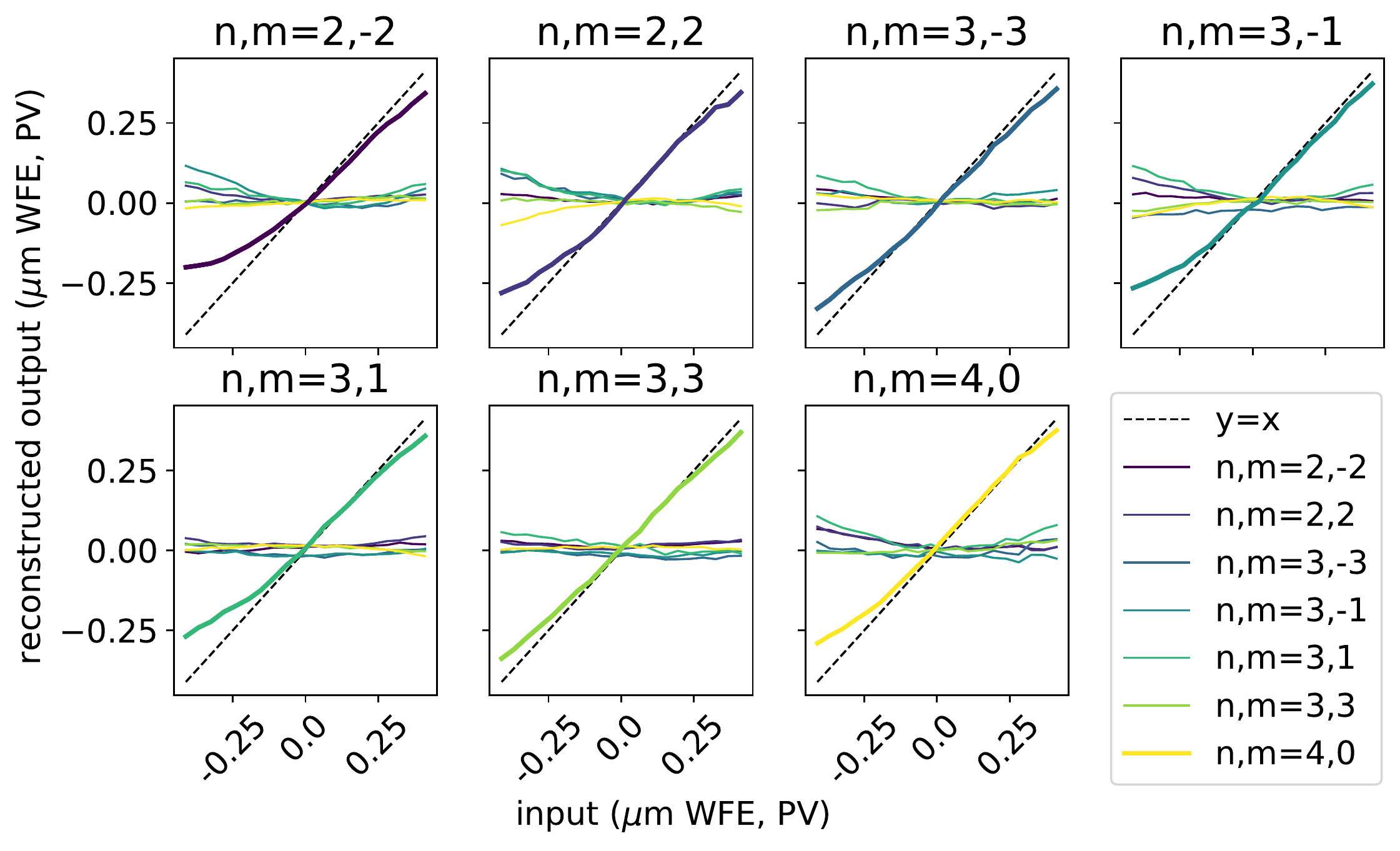}
    \caption[LOWFS linearity for Zernike modes beyond tip/tilt/focus]{LOWFS linearity for Zernike modes beyond tip/tilt/focus, labeled by their n and m indices.}
    \label{fig: zernike_linearity}
\end{figure}

\begin{figure}[!h]
    \centering
    \includegraphics[width=1.0\textwidth]{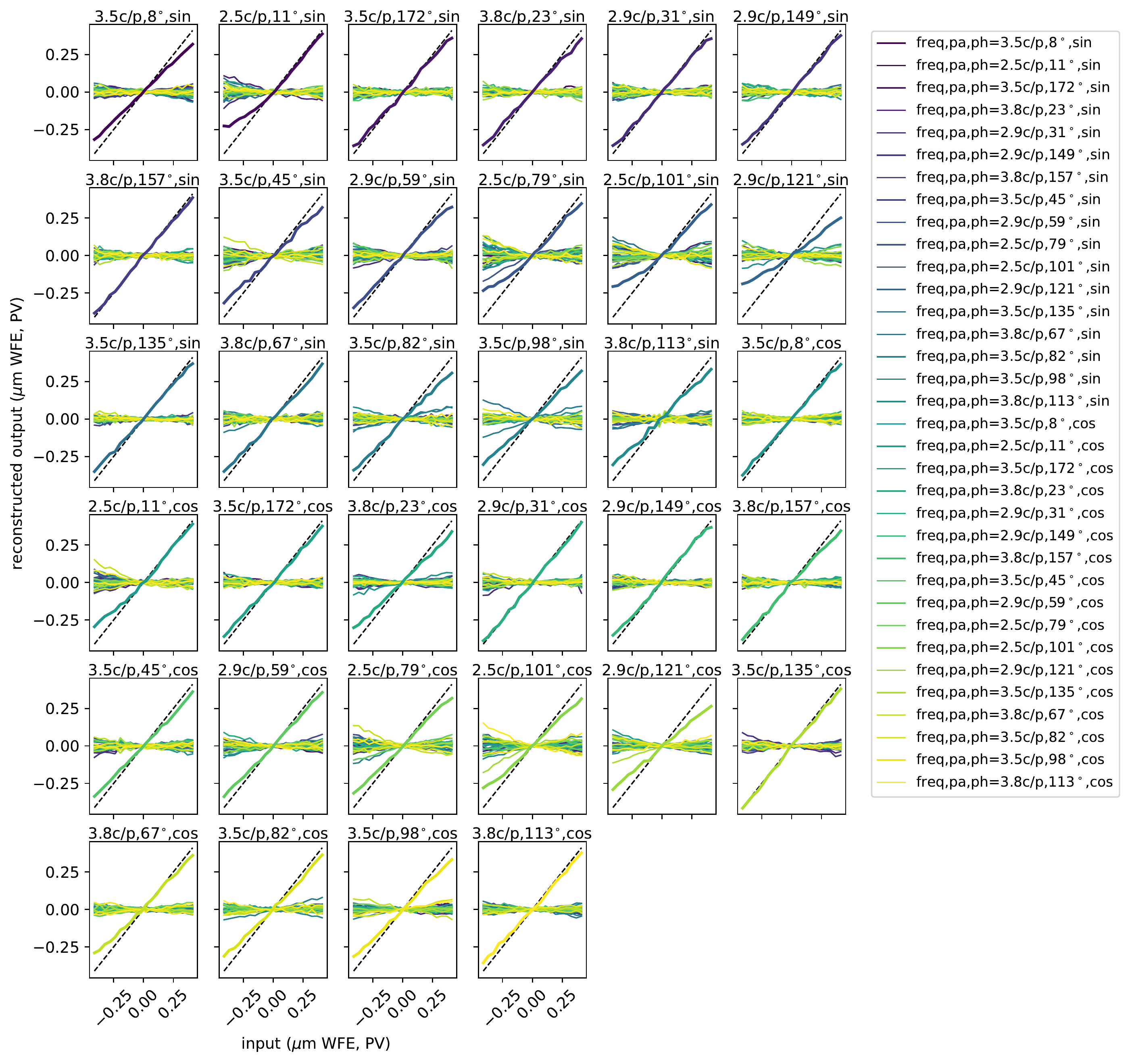}
    \caption[LOWFS linearity for low order Fourier modes]{LOWFS linearity for Fourier modes between 2 and 4 c/p, indicated by the amplitude, position angle, and relative phase of their spatial frequency (labeled ``freq,pa,ph,'' respectively.) The relative phase is reconstructed from a pair of sine and cosine modes for a given spatial frequency, as labeled. All panels show the same x and y scale of $\pm$0.45 $\mu$m WFE.}
    \label{fig: lowfs_linearity}
\end{figure}

We apply a similar MVM-based approach to generating a IM and CM for low order modes as described in \S\ref{sec: howfs_recon} for high order modes. During the IM generation, instead of applying binary masks around individual sine spots, a binary mask in the $I_-$ plane uses pixel values only within a 5 $\lambda/D$ radius---the limiting DH inner working angle (IWA) for this system\cite{fast_spie18}---of the optical axis. After testing and development, we resolved to implement closed-loop LOWFS performance for three different modal groups: (1) tip/tilt/focus (TTF), (2) the next five Zernike modes, and (3) low order Fourier modes between 2 and 4 c/p with modal groups (1) and (2) de-projected from this basis. The linearity curves, showing reconstructed output wavefront error (WFE) vs. input WFE amplitude for all modes applied and measured, are in Figures \ref{fig: ttf_linearity} - \ref{fig: lowfs_linearity}.

As shown, these modal groups form an excellent basis set to enable low order control, with a clear linear regime defined for each mode and minimal cross-talk between other modes in this regime. Fig. \ref{fig: ttf_linearity} shows a tip/tilt capture range of $\sim$0.2 $\mu$m PV corresponds to $\sim$0.3 $\lambda/D\approx$10 mas for a 8m telescope at $\lambda$=1.6 $\mu$m, consistent with simulations from Ref. \citenum{fast_spie20} and the need for a diffraction-limited AO residual wavefront (i.e, Strehl ratio $\gtrsim$20\%) to close the FAST loop. For each modal group, the input ``poke'' amplitude used to generate the IM and SVD cutoff used to generate the CM are separately optimized to produce optimal linearity and minimal cross-talk. It is for this reason that we placed focus in the same modal group as tip/tilt rather than with the other Zernike modes: a single SVD cutoff value produced good linearity for all Zernikes other than focus, where as the same tip/tilt SVD cutoff preserved good linearity for focus. We also confirmed that if all modes were combined in the same IM and CM that cross-talk between any mode and all the other modes was still negligible within the linear range of that mode (although with a single SVD cuttoff for all modal groups, linearity appeared qualitatively worse than as shown in Figures \ref{fig: ttf_linearity} - \ref{fig: lowfs_linearity}, which is why we chose separate these groups into different IMs and CMs). The robustness of our choice to use separate SVD cutoffs and IM amplitudes for each modal group is also further supported by closed-loop results in \S\ref{sec: closed_loop} (i.e., if cross-talk between modal groups remained significant, the loops would not close and/or go unstable).

Interestingly, Fig. \ref{fig: ttf_linearity} shows some conceptual discrepancies from TTF linearity simulations in Ref. \citenum{fast_spie20}, including a smaller capture range for tip and for tilt and cross talk in focus only from tip but not from tilt. As we have not observed this behavior in simulation and we have already optically aligned the system for these modes, we conclude (by process of elimination) that these behaviors are due to FPM fabrication defects. Ref. \citenum{fast_spie20} did not analyze SCC/LOWFS modes beyond TTF, so we are not able to compare simulations to Figures \ref{fig: zernike_linearity} and \ref{fig: lowfs_linearity}, but in general the strong linearity and low cross talk for each mode indicates the potential for efficient closed-loop control, which is the main focus of this paper.

\section{CLOSED-LOOP OPERATIONS}
\label{sec: closed_loop}
This section is split into three, first presenting results and analysis from a generating a static DH (\S\ref{sec: static}), second from closing the FAST LOWFS and high order WFS (HOWFS) loops for on-air quasi-static errors within the SEAL testbed enclosure (\S\ref{sec: quasi_stat}), and lastly in closing the LOWFS and HOWFS loops on simulated AO residual turbulence (\S\ref{sec: dynamic}). The latter is an important milestone for focal plane wavefront sensing technologies, demonstrating the potential of FAST to be implemented as a second stage AO WFS. In particular, the ability to enable low and high order control is critical, as the latter is dependent on the former to enable optimal performance gains\cite{guyon18}. Common to all subsections, our current Python-based real-time code runs at approximately 50 Hz with 1 frame of system latency, set by a user-defined 20 ms pause at each iteration to enable FAST images (at the end of the iteration) to see the effect of DM commands applied (at the beginning of the iteration); we found that pauses smaller than this amount could prevent this sequential dependence from functioning properly, although we do not expect this to remain an issue after a planned future migration to a parallelized framework (see \S\ref{sec: setup}). 

Common to \S\ref{sec: quasi_stat} and \S\ref{sec: dynamic}, we record HOWFS and LOWFS IMs and CMs (as described in \S\ref{sec: howfs_recon} and \S\ref{sec: lowfs}, respectively) immediately before running an open and closed loop sequences, minimizing the impact of time-dependent control errors. LOWFS corrections are calibrated and applied on the 11x11 ALPAO DM (hereafter the low order DM, or LODM), while HOWFS corrections use the 34x34 MEMS (here after the high order DM, or HODM). Closed-loop target SCC images use differential images (with respect to a static DH image taken just before the sequence begins) dotted with the CM (i.e., Equation \ref{eq: dmc}) for each modal group to generate DM commands (i.e., instead of absolute speckle correction, preventing achievable closed-loop contrasts deeper than the static DH but enabling faster convergence to closed-loop residual levels). For all sections we use a leaky integrator controller with the leaks and gains tuned separately by hand (and separately for each section) for each modal group (but with a fixed value for each parameter across a given modal group), balancing rejection of temporal disturbances (which prefers higher gains and leaks) with noise propagation and loop stability (which prefers lower gains and leaks)\cite{gendron94}.
\subsection{Static Errors}
\label{sec: static}
\begin{figure}[!h]
    \centering
    \includegraphics[width=1.0\textwidth]{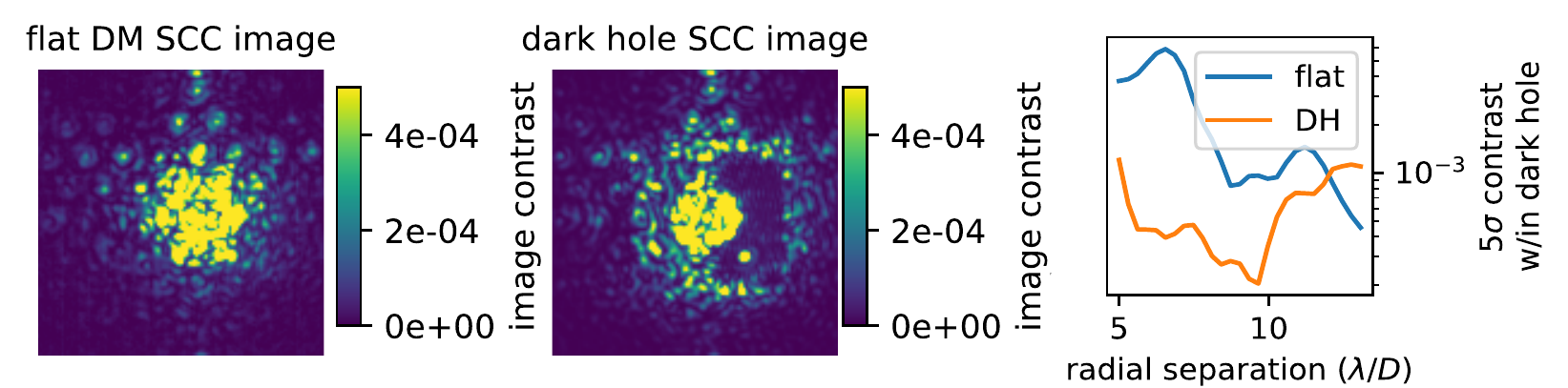}
    \caption[Static dark hole generation]{Speckle correction/subtraction results of quasi-static aberrations on SEAL. Left: FAST image before DM correction. Middle: The same image after DM correction, generating a dark hole on the right side of the star with the planet visible to the lower right of the star. Right: 5$\sigma$ contrast curves of the SCC image within the dark hole (DH) region (masking the planet signal) for before (``flat'') and after (``DH'') static correction, showing up to a 10x improvement.}
    \label{fig: stat_imas_and_ccurves}
\end{figure}
Fig. \ref{fig: stat_imas_and_ccurves} shows images and contrast curves of speckle correction by DM control of static aberrations within the SEAL system. The off-axis planet source (to the lower right of the star) is visible in the middle ``dark hole SCC image'' panel. Due to the static nature of this initial DH, two control approaches produce similar results: (1) using a low gain integral controller (typically below 0.01) to ensure loop convergence over a large number of iterations (converging typically after a few hundred frames) or (2) using a higher gain but limiting the number of iterations, beyond which the loop diverges due to detector and photon noise propagation into the reconstructed DM commands (e.g., a gain of 0.1 fixed at 100 iterations typically produces similar results as option 1). Contrast curves in the right panel of Fig. \ref{fig: stat_imas_and_ccurves}, calibrated to the peak off-axis stellar flux as described in \S\ref{sec: appendix_calibration}, compute five times the standard deviation within a given radial annulus of the SCC image (masking the planet signal to prevent bias in computing contrast gains) that is also within the half DH region on the right hand side of the image, showing up to a 10x contrast gain depending on separation for the shown 20 ms exposure time. In the right panel of Fig. \ref{fig: stat_imas_and_ccurves} it is also apparent that although we defined the DH control region out to 12 $\lambda/D$ along the x direction (\S\ref{sec: howfs_recon}), contrast gains are negligible beyond $\sim$10 $\lambda/D$ due to a trade-off we found between SVD cutoff and MEMS stroke limitations: a lower SVD cutoff (i.e., including more eigen modes in the HOWFS CM) produced a deeper DH between 10 and 12 $\lambda/D$ but also required $\sim$2x more stroke (which cannot be compensated by the higher-stroke ALPAO device due to the spatial frequencies being considered) than a higher SVD cutoff, which produced a DH as deep out to 10 $\lambda/D$ but not as deep between 10 and 12 $\lambda/D$. Given the subsequent need to use additional MEMS stroke to apply and then correct AO residual turbulence (\S\ref{sec: dynamic}), we chose to proceed with the higher SVD cutoff option.
\subsection{Quasi-static Errors}
\label{sec: quasi_stat}

\begin{figure}[!h]
    \centering
    %\begin{subfigure}[c]{1.0\textwidth}
    %    \includegraphics[width=1.0\textwidth]{cl_air_one_mode.pdf}
    %    \caption{On-air real-time correction of four individual modes.}
    %\end{subfigure}
    %\begin{subfigure}[c]{1.0\textwidth}
    %    \includegraphics[width=1.0\textwidth]{cl_air.pdf}
    %    \caption{On-air real-time correction of all modes for all four modal groups.}
    %\end{subfigure}
    \includegraphics[width=1.0\textwidth]{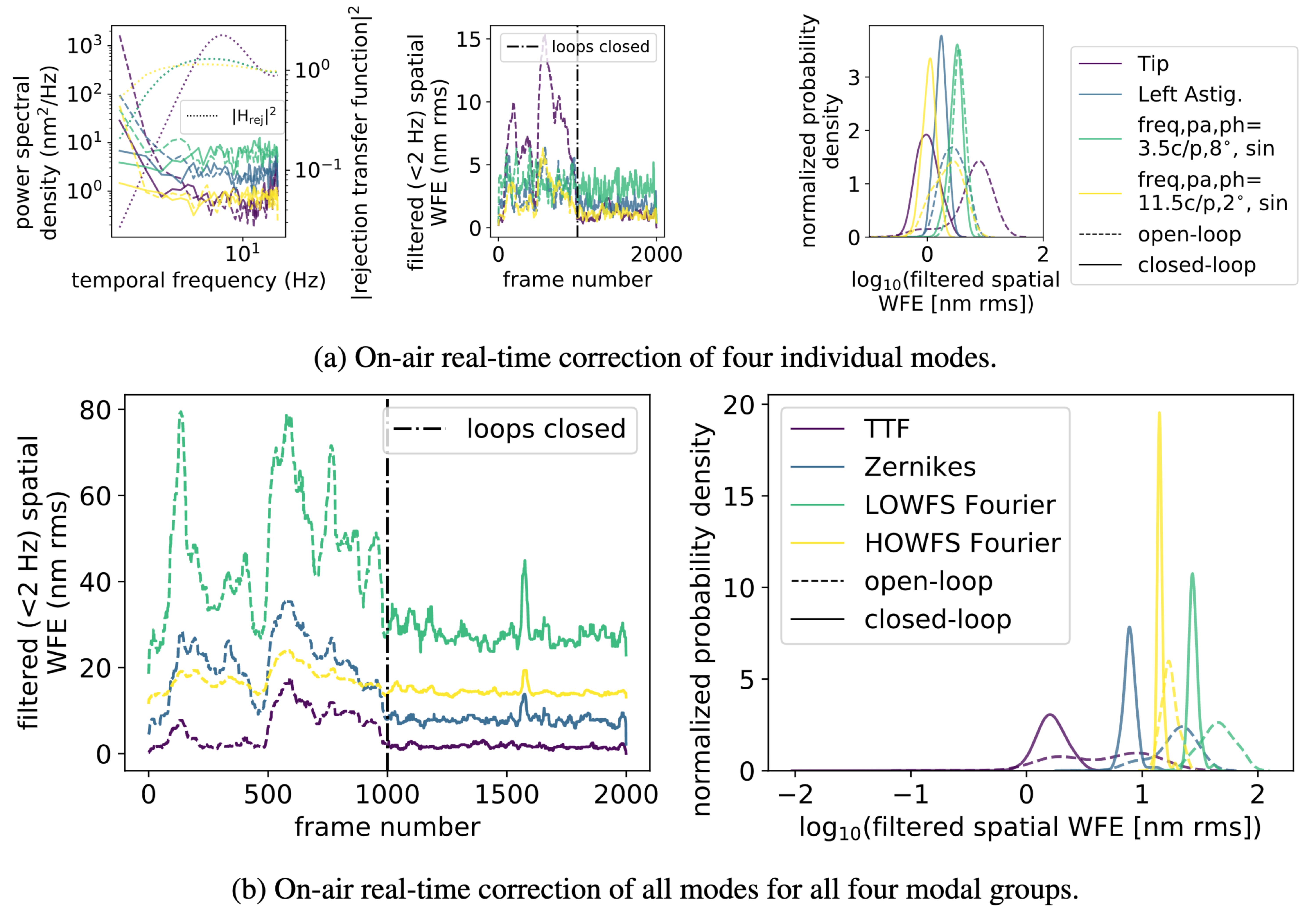}
    \caption[Quasi-static dark hole stabilization]{SEAL temporal stability before and after closing the loop on only quasi-static turbulence. Panel a shows the statistics for a single modal within each of the four modal groups (with each group indicated by a different color; the color scheme is consistent across all panels in this figure), while panel b shows the cumulative statistics for all modes within a given modal group. Temporal power spectral density plots on the left panel side of panel a motivate the subsequent display of low-pass filtered wavefront statistics, where the choice of temporal frequency cutoff (2 Hz) is the 0 dB bandwidth for tip. Kernel density distributions are computed using the \texttt{seaborn.kdeplot} package\cite{Waskom2021}. All panels clearly illustrate a WFE reduction and stabilization for all modes.}
    \label{fig: static_turb}
\end{figure}
Fig. \ref{fig: static_turb} shows the temporal effects of closing the loop on air, stabilizing any quasi-static effects such as air flow through the enclosure. The histograms in Fig. \ref{fig: static_turb} clearly show a gain in WFE for closed vs. open loop control for all modal groups. The median TTF open- and closed-loop WFE are 6.2 and 1.6 nm rms, respectively, which for the latter corresponds to 0.003 $\lambda/D$ for tip/tilt, illustrating the highly stabilized environment the SEAL granite testbed and enclosure enables. Although this nm level of on-air stability is perhaps unrealistic for an on-sky ground-based environment due to realistic quasi-static effects (e.g., mechanical effects from telescope tracking, changing thermal environment, instrument flexure for Cassegrain instruments, etc.), it sets the stage for similar gains to be made on AO residual turbulence correction, discussed next.
\subsection{Dynamic AO Residual Errors}
\label{sec: dynamic}
\begin{figure}[!h]
    \centering
    \includegraphics[width=1.0\textwidth]{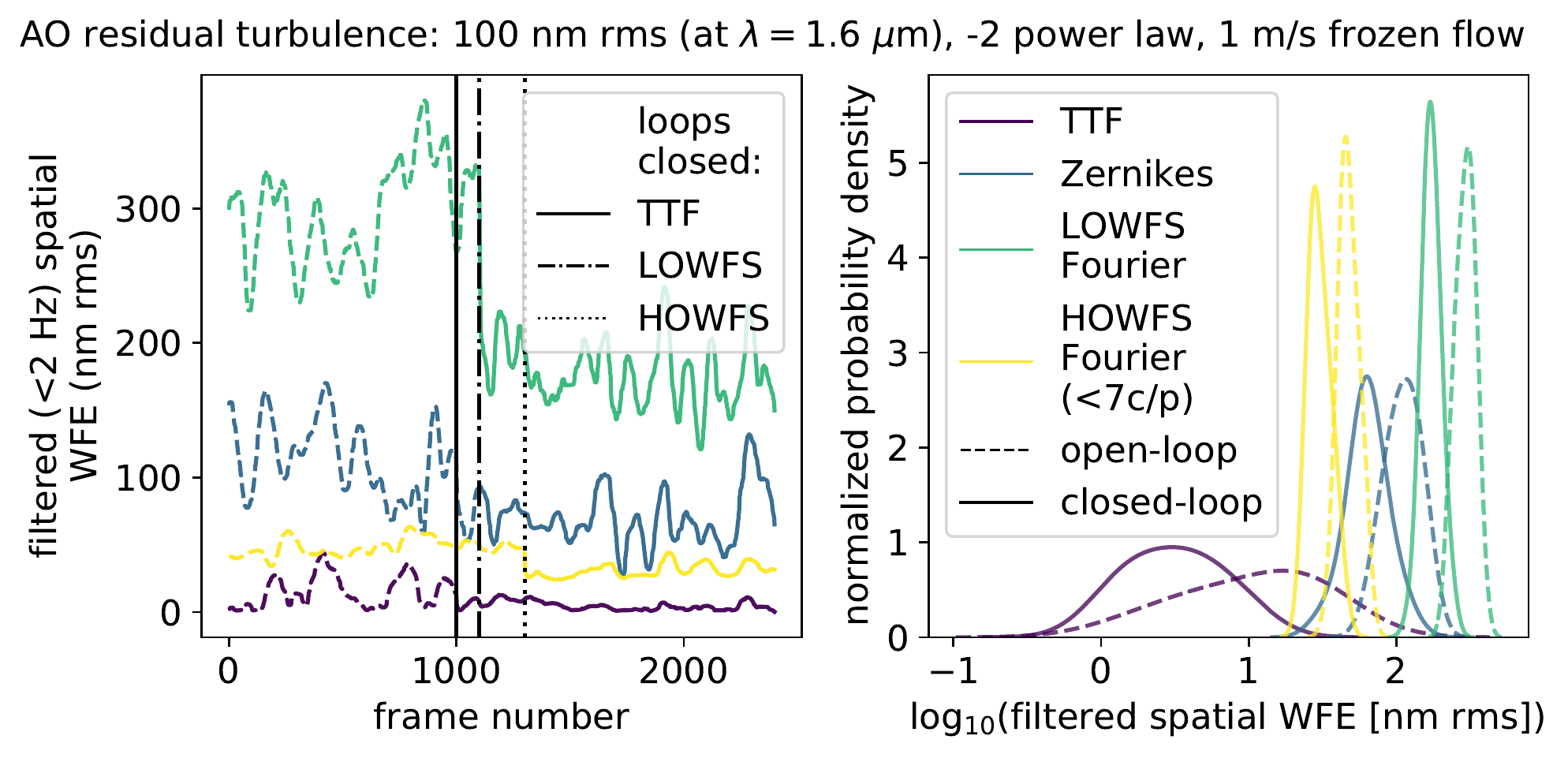}
    \caption[FAST AO residual corrections, part 1]{Time series (left) and corresponding kernel density distributions (right) of open and closed loop operations of AO residual turbulence, using the same temporal filter based on the analysis in Fig. \ref{fig: static_turb}. The same colors in the right panel legend also apply to the left panel, noting that high order Fourier modes are only shown for spatial frequencies between 5 and 7 c/p. Atmospheric parameters are shown in the title; the loop speed is approximately 50 Hz with 1 frame delay. Gains and leaks for a leaky integrator controller are separately tuned for each modal group. Loop closures occur at three different times for tip/tilt/focus (TTF), other low order WFS (LOWFS) modes (including Zernike and LOWFS Fourier modes), and high order WFS (HOWFS) modes, indicated by the three different vertical lines on the left panel. Open and closed loop distributions in the right panel are computed with data in the left panel to the left of the solid line (TTF loops closed) and to the right of the dotted line (HOWFS loops closed), respectively.}
    \label{fig: cl_turb}
\end{figure}

\begin{figure}[!h]
    \centering
    \includegraphics[width=1.0\textwidth]{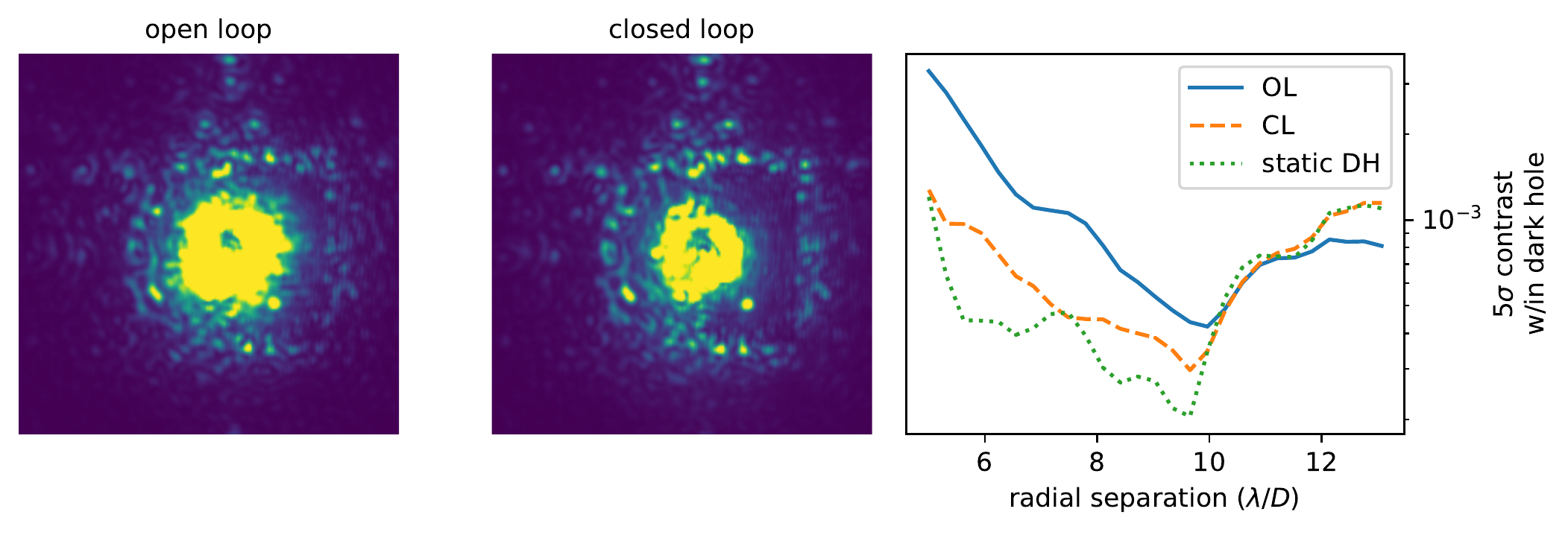}
    \caption[FAST AO residual corrections, part 2]{Stacked FAST images (on the same relative contrast scale) and corresponding contrast curves for open and closed loop (OL and CL, respectively) control of simulated AO residual turbulence (using the same frame indices to define the OL and CL regimes as in the right panel of Fig. \ref{fig: cl_turb}), also showing the same ``static DH'' contrast curve from Fig. \ref{fig: stat_imas_and_ccurves} for reference. The same planet source is present in the lower right side of the DH images. Closed-loop correction of AO residual turbulence is demonstrated to improve contrast by up to a factor 5 relative to no residual atmospheric correction (OL).}
    \label{fig: dynamic_imas_and_ccurves}
\end{figure}

In this section we present an important milestone for focal plane wavefront control technologies, demonstrating FAST correction of AO residual speckles in quasi-real-time for both low and high order modes. An evolving full power law AO residual wavefront (i.e., including low order modes) is applied on the MEMS in open loop and then corrected by the FAST LOWFS and HOWFS in closed-loop. AO residual WFE is normalized to 100 nm rms between 0 and 16 c/p assuming $\lambda=1.6\mu$m with a -2 spatial power spectral density (PSD) power law. A 10m diameter telescope, 1 m/s ground layer ``frozen flow'' wind speed is assumed.\footnote{As discussed at the beginning of \S\ref{sec: closed_loop}, our Python real-time code limits operational frame rate to 50 Hz with a 1 frame delay. This limited loop speed is the main reason why a 1 m/s wind speed is used, although a more realistic 1 ms atmospheric lag, as motivated in Ref. \citenum{fast_spie18}, is applied at the time of DM correction. Regardless, future more-optimized real-time code running at kHz speeds will enable closed-loop correction on more typical wind speeds of order 10 m/s.}
With this setup implemented, Fig. \ref{fig: cl_turb} shows the time series and WFE distributions and Fig. \ref{fig: dynamic_imas_and_ccurves} shows long exposure SCC images and contrast curves (computed analogously to Fig. \ref{fig: stat_imas_and_ccurves} as described in \S\ref{sec: static}) for open and closed-loop FAST correction of evolving AO residuals.

Fig. \ref{fig: dynamic_imas_and_ccurves} shows up to a 5x contrast gain comparing open and closed loop correction of AO residual turbulence, illustrating the benefit of closed loop operations at all spatial frequencies. Fig. \ref{fig: cl_turb} shows a 1.5x HOWFS median WFE gain (46 vs. 30 nm rms for open vs. closed-loop HOWFS WFE, respectively, for spatial frequencies between 5.5 and 7 c/p), which is consistent with the 2.3x median OL vs. CL contrast gains between 5.5 and 7 $\lambda/D$  in Fig. \ref{fig: dynamic_imas_and_ccurves} (i.e., contrast gains should scale as the square of the WFE gains). Fig. \ref{fig: cl_turb} also illustrates the sequential process of closing loops: in contrast to on-air control in Fig. \ref{fig: static_turb}, here we found that all loops cannot be closed at once, requiring initial TTF loop closure, then LOWFS (including Zernike and Fourier modes simultaneously), and finally HOWFS Fourier modes. When we tried to close all modes simultaneously, or, e.g., HOWFS modes first, the loop would go unstable shortly thereafter. This is to be expected: LOWFS modes must be closed and stabilized first (i.e., producing a stable stellar core on the FAST FPM, which is needed for HOWFS measurements), as we predicted in Ref. \citenum{fast_spie20}. The number of frames needed between closing the TTF, LOWFS, and HOWFS loops for the different modal groups was determined by applying AO residuals on the MEMS with only the modes from a given group applied and then determining by eye how long it took for the loop to converge; we did not further explore flexibility in this parameter space. Contrast improvement in Fig. \ref{fig: dynamic_imas_and_ccurves} decreases with separation, however, suggesting that loop control is not yet optimized. In Fig. \ref{fig: cl_turb}, we only plot HOWFS telemetry for spatial frequencies below 7 c/p, as including these higher order modes ``dilutes'' the relative gain from these plotted HOWFS modes. By applying a single gain and leak (0.3 and 0.8, respectively) to all HOWFS modes (i.e., as described in \S\ref{sec: scc_dm}), there is an inherent non-optimal trade-off between temporal rejection at lower spatial frequencies vs. noise amplification at higher spatial frequencies\cite{gendron94}; our chosen scalar gain and leak values gave better rejection at lower spatial frequencies and more noise amplification at higher ones, consistent with the smaller contrast gains at larger separations in Fig. \ref{fig: dynamic_imas_and_ccurves}. This clearly motivates the need to implement more advanced control, such as modal gain optimization.\cite{fast_spie18} Further comparing the closed-loop performance to static DH limits (dashed orange vs. dotted green curves), it is clear that across most spatial frequencies there are additional gains to be made with real-time correction of AO residuals. We discuss such further areas for contrast improvement in \S\ref{sec: improvement}.

Regardless of additional gains to be made, Fig. \ref{fig: cl_turb} and particularly Fig. \ref{fig: dynamic_imas_and_ccurves} represent an important milestone in using real-time focal plane wavefront control to boost the achievable final contrasts with exoplanet imaging instruments. To be clear, this is not the first demonstration of real-time focal plane wavefront control; Ref. \citenum{scexao_ldfc} have demonstrated on-sky linear dark field control at 20 Hz, but with negligible contrast gains due to a noisy detector and insufficient bandwidth for control of residual atmospheric speckles. In this paper, however, we have demonstrated a clear raw contrast gain of up to 5x for simulated AO residuals in real-time. The implications of this result scales to an up to 5x final contrast gain, supported by the work from Ref. \citenum{bailey16} (correlating AO performance to achievable raw contrast and raw contrast to final contrast). In other words, this demonstration of 5x raw contrast gain with FAST translates to boosting post-processed contrast by 5x with angular differential imaging\cite{adi} (even in second generation high contrast imaging instruments, spectral differential imaging\cite{sdi} yields negligible contrast gains to due significant instrument chromaticity\cite{fast_phd}).
\section{KEY AREAS FOR CONTRAST IMPROVEMENT}
\label{sec: improvement}
Perhaps the most significant area for further improvement is the optimal choice of LOWFS modal bases to reduce the impact of non-linearities on the achievable closed-loop WFE.  The choice of Fourier modes for LOWFS control is perhaps a non-intuitive one; applying DM Fourier modes with low enough spatial frequencies ($<3$ c/p for this TG FPM) generates sine spots that are simply blocked by the FPM and redirected to the off-axis FAST pupil, rather than generating sine spots in the coronagraphic image. We could instead consider higher order Zernike modes; however, these increasingly become a non-natural basis with increasing radial order, as entrance pupil wavefront spatial frequencies corresponding to focal plane separations beyond the TG FPM core are not sent to the off-axis Lyot stop pupil, instead overlapping (i.e., increasing non-linear cross-talk) with the FAST HOWFS Fourier modal group. Instead, we first tried applying a basis of higher order spatially-filtered ($<$ 3 c/p, using an algorithmic Butterworth Fourier filtering mask) Zernike modes, but at the cost of higher order Zernike modes becoming increasingly non-linear due to the loss of information by a low-pass filter.\footnote{Alternatively, considering the PSFs of Zernike modes, PSF structure deviating from an Airy pattern increases in radial extent as a function of Zernike mode radial order when amplitudes are otherwise fixed, and so restricting a smaller region of the PSF core to use for wavefront reconstruction causes increased non-orthogonality vs. Zernike mode order.} Thus, a LOWFS Fourier modal basis is instead a natural choice, restricting diffracted light to within the FPM IWA. However, for reasons we do not yet understand, extending this Fourier modal basis to spatial frequencies \textit{below} 2 c/p also did not produce good linearity, while low order Zernike modes instead did, which motivates our choice of both modal bases (the Zernike mode cutoff was determined based on the mode above which non-linearities would significantly increase if spatially filtered to less than 2 c/p). Other bases should be further explored (e.g., KL modes, disk harmonics, etc.) that can span the full range of accessible LOWFS spatial frequencies (or perhaps in combination with Fourier and/or Zernike modes, as was done in this manuscript) to optimally ``linearize'' FAST LOWFS performance. Furthermore, such additional bases should be explored specifically in the context of realistic obstructed pupil geometries for 8m-class telescopes (e.g., using annular Zernike polynomials), as the orthogonality of bases used in this paper may be less robust in the presence of such pupil geometries.

Higher quality of fabricated FAST FPMs will also enable further gains. In Ref. \citenum{fast_spie21} we modeled that our current TG FPM has around 300 nm rms of phase errors; In Ref. \citenum{fast_spie18} we showed that getting such errors below 100 nm rms would enable TG fringe visibilities at 10 $\lambda/D$ up to 15\% (about 10x greater than what we are getting now). Such a higher quality mask would improve measurement sensitivity to photon noise (i.e., lowering closed-loop WFE for a fixed number of input photons) and also non-linearities. We noticed the effects of the latter due to fabrication defects on our FPM, where some specific LOWFS Fourier modes were intentionally not controlled due their empirically measured bad linearities (while other modes at the same spatial frequency but different position angle remained sufficiently linear). We have already demonstrated that it is realistic to fabricate FAST FPMs at the nm rms level,\cite{fast_ao4elt,fu2021} so it is reasonable to expect that future FAST FPMs can improve on this front.

%- note that LOWFS non-linearities and in general WFE improving means that HOWFS non-linearities will also improve (optical gain effect). 
Additional algorithmic optimization procedures can also enable further contrast gains. Approaches to improving HOWFS linearities include varying the Fourier modal amplitude with spatial frequency so that IM spot intensities optimally match the power law of AO residual aberrations,  optimizing the binary mask radius used to isolate sine spots during the HOWFS IM calibration (currently fixed at 1.3 $\lambda/D$; \S\ref{sec: howfs_recon}), and considering different modal groups within the HOWFS control region to control separately with different SVD cutoffs and CMs (e.g., as has already been done for LOWFS TTF, Zernike, and Fourier modes in \S\ref{sec: lowfs}). FAST modal gain optimization\cite{fast_spie18}, both for LOWFS and HOWFS control is also a key area to develop that we have already been working on with SEAL\cite{fast_spie21}. However, this approach requires characterizing the rejection transfer function (RTF\cite{correia18}; i.e., which is defined for a given mode relative to it's open and closed-loop temporal PSDs, or PSD$_\mathrm{OL}$ and PSD$_\mathrm{CL}$, respectively, as: RTF $\equiv\sqrt{\mathrm{PSD}_\mathrm{OL}/\mathrm{PSD}_\mathrm{CL}}$), a typically linear assumption for which FAST non-linearities may pose limitations to it's success. FAST predictive wavefront control, which we are also beginning to implement\cite{vankooten21a}, is also a critical area to develop, as the inherent \textonehalf\; frame delay only recoverable by prediction is much larger at the slower 100 Hz speeds FAST will be limited to on 5th magnitude stars with an 8m telescope\cite{fast18} (note that the data acquired in this paper, collecting hundreds of photons per pixel per frame, is more akin to a 0th magnitude star); data-driven predictive control methods (e.g., Ref. \citenum{vankooten21b}) can also be used to ``learn'' FAST non-linearies, also further improving achievable gains. 
%- higher loop speeds (noting that at the moment PWFC on lag doesn't gain much due to the slow frame rate) to enable higher wind speeds
%- PWFC to enable lag compensation at higher wind speeds and higher loop speeds AND non-linearity compensation in general
%
%
\section{CONCLUSION}
\label{sec: conclusion}
Exoplanet imaging speckle subtraction via focal plane wavefront control is a promising technology that has the potential to enable current and future instruments to detect and characterize lower mass, closer in, and/or older exoplanetary systems than is currently possible. However, one issue thus far in making such technology operational on-sky has been achievable temporal bandwidth, due to continual speckle evolution from atmospheric and quasi-static effects. We have developed a technique to mitigate this bandwidth issue, called the Fast Atmospehric Self-coherent camera Technique (FAST), for which in this manuscript we present laboratory validation on the Santa Cruz Extreme AO Laboratory (SEAL) testbed. Our main findings are as follows:
\begin{enumerate}
    \item We tested FAST quasi-static DH generation and closed-loop temporal stability, demonstrating 5$\sigma$ contrasts within the DH of around 5e-4 (Fig. \ref{fig: stat_imas_and_ccurves}) and nm rms-level temporal stabilities on-air (Fig. \ref{fig: static_turb}).
    \item We tested FAST correction of AO residual speckles, demonstrating a clear gain of such a real-time ``second stage AO'' capability using a focal plane wavefront sensor, showing up to a factor of 5x contrast gain, depending on radial separation (Fig. \ref{fig: dynamic_imas_and_ccurves}).
\end{enumerate}
These testing results clearly illustrate the benefit of running real-time focal plane wavefront control technology. However, the developments presented here are by no means comprehensive, instead opening many new avenues of real-time focal plane wavefront control approaches to explore and optimize, some of which we discuss in \S\ref{sec: improvement}. In future manuscripts we also plan to investigate FAST SEAL testing of many related additional topics, including increasing spectral bandwidth capabilities
(for which fringe smearing due to PSF magnification with wavelength can be mitigated with the multi-reference SCC\cite{mrscc} adapted specifically for FAST\cite{fast_phd})
, optimizing mid-to-high order temporal bandwidth (including predictive control), real-time integration with a first stage AO correction, and optimizing control for changing atmospheric conditions and corresponding first stage AO performance.
\section*{Acknowledgments}
We gratefully acknowledge research support of the University of California Observatories for funding this research. We thank Maaike van Kooten, Jules Fowler, Rachel Bowens-Rubin, Dominic Sanchez, Renate Kupke, and Phil Hinz for comments, suggestions, and discussions that have contributed to this manuscript. We thank the reviewers for their helpful feedback and suggestions that improved this manuscript.
\appendix

\section{Summary of the Self-Coherent Camera}
\label{sec: scc_sum}
The SCC, developed by Ref. \citenum{scc_orig}, \citenum{galicher_scc}, and others, is a method designed to use the coronagraphic science image as both a WFS to control an upstream DM and enable CDI-based post-processing. We review the basics of classical (i.e., not FAST-specific) SCC image formation and processing (\ref{sec: scc_im}) and DM control (\ref{sec: scc_dm}). SCC CDI techniques are not summarized due to the scope of this paper.
\subsection{Image Formation and Processing}
\label{sec: scc_im}
\begin{figure}[!h]
    \centering
    %\begin{subfigure}[c]{1.0\textwidth}
    %\centering
    %\includegraphics[width=0.7\textwidth]{scc_geometric.pdf}
    %\caption{Geometric optics SCC illustration. Blue lenses represent focusing/collimating optics; all other components are as labeled in a pupil (entrance pupil, Lyot stop) or focal (FPM, detector) plane.}
    %\end{subfigure}
    %\begin{subfigure}[c]{1.0\textwidth}
    %\includegraphics[width=1.0\textwidth]{scc.pdf}
    %\caption{Fourier optics SCC illustration.}
    %\end{subfigure}
    \includegraphics[width=1.0\textwidth]{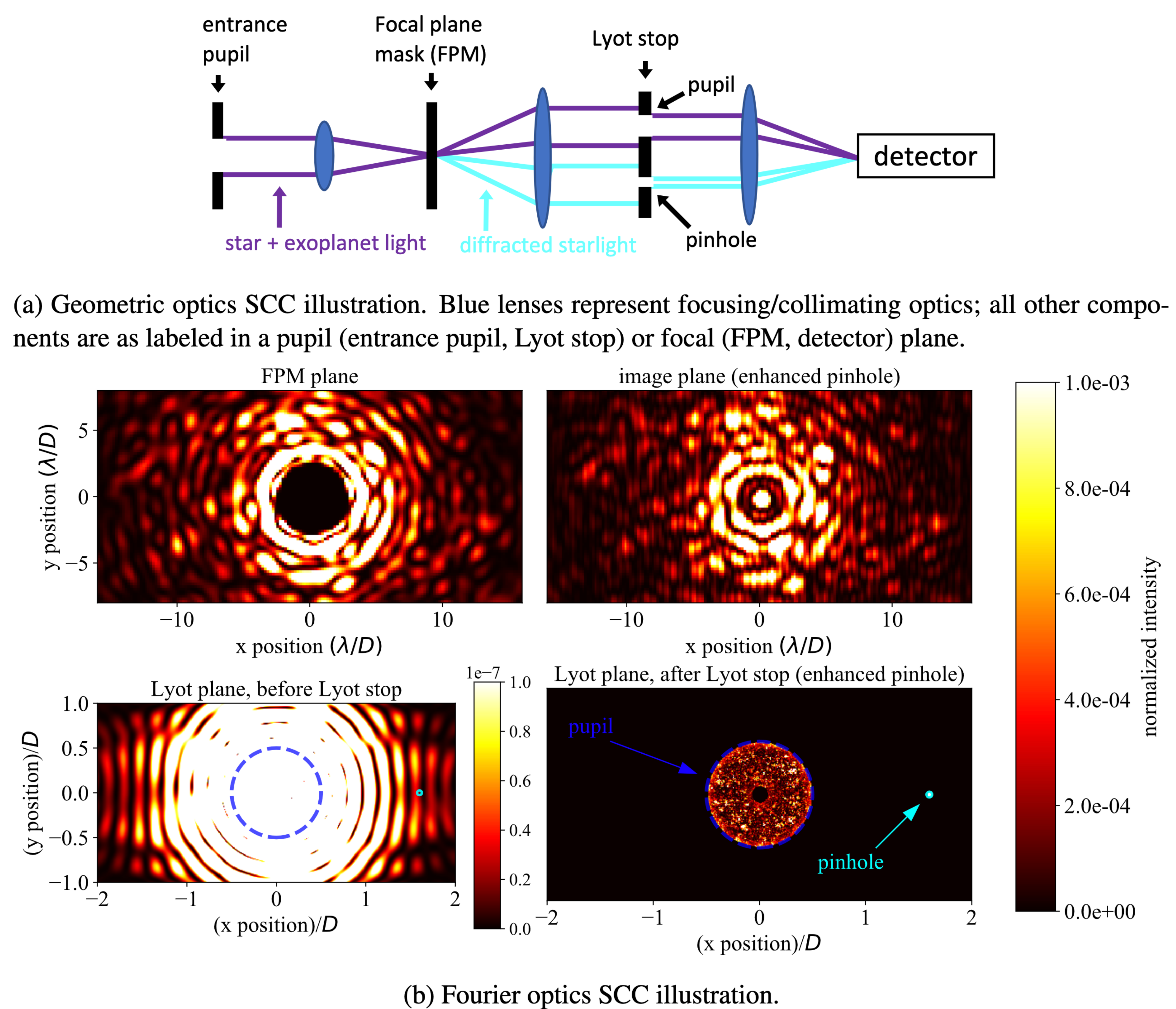}
    \caption[Illustration of SCC image formation]{
    An simulated illustration of SCC image formation due to the geometric (a) and diffractive (b) properties of a focal plane mask (FPM), adapted from Ref. \citenum{fast18}. We describe the four sub-panels in panel b. \uline{Upper left}: in an intermediate focal plane, a FPM blocks the on-axis star while transmitting remaining diffracted starlight and off-axis exoplanet light. \uline{Lower left}: in the downstream pupil plane, starlight is diffracted outside the footprint of the primary mirror if no FPM was deployed (illustrated by the dashed blue circle); such diffraction effects occur for any FPM with sub-$\lambda/D$ spatial scales, in this case encoded by the hard edge of an amplitude ``dot'' FPM. \uline{Lower right}: a Lyot stop blocks most of this diffracted light outside the pupil footprint, except for a small off-axis pinhole (whose cyan aperture footprint is also illustrated in the lower left panel) that is transmitted downstream. \uline{Upper right}: The Lyot stop pinhole and pupil apertures interfere in the coronagraphic image to form fringes on sub-$\lambda/D$ spatial scales, overlaid on the existing speckle pattern (with spatial scales larger than $\lambda/D$) from the main Lyot stop pupil aperture. The electric field intensity in the pinhole aperture used to simulate the lower and upper right panels is (un-physically) enhanced by a factor of 2.5$\times$10$^5$ for the purposes of this illustration.
    }
    \label{fig: scc}
\end{figure}

\begin{figure}[!h]
    \centering
    \includegraphics[width=1.0\textwidth]{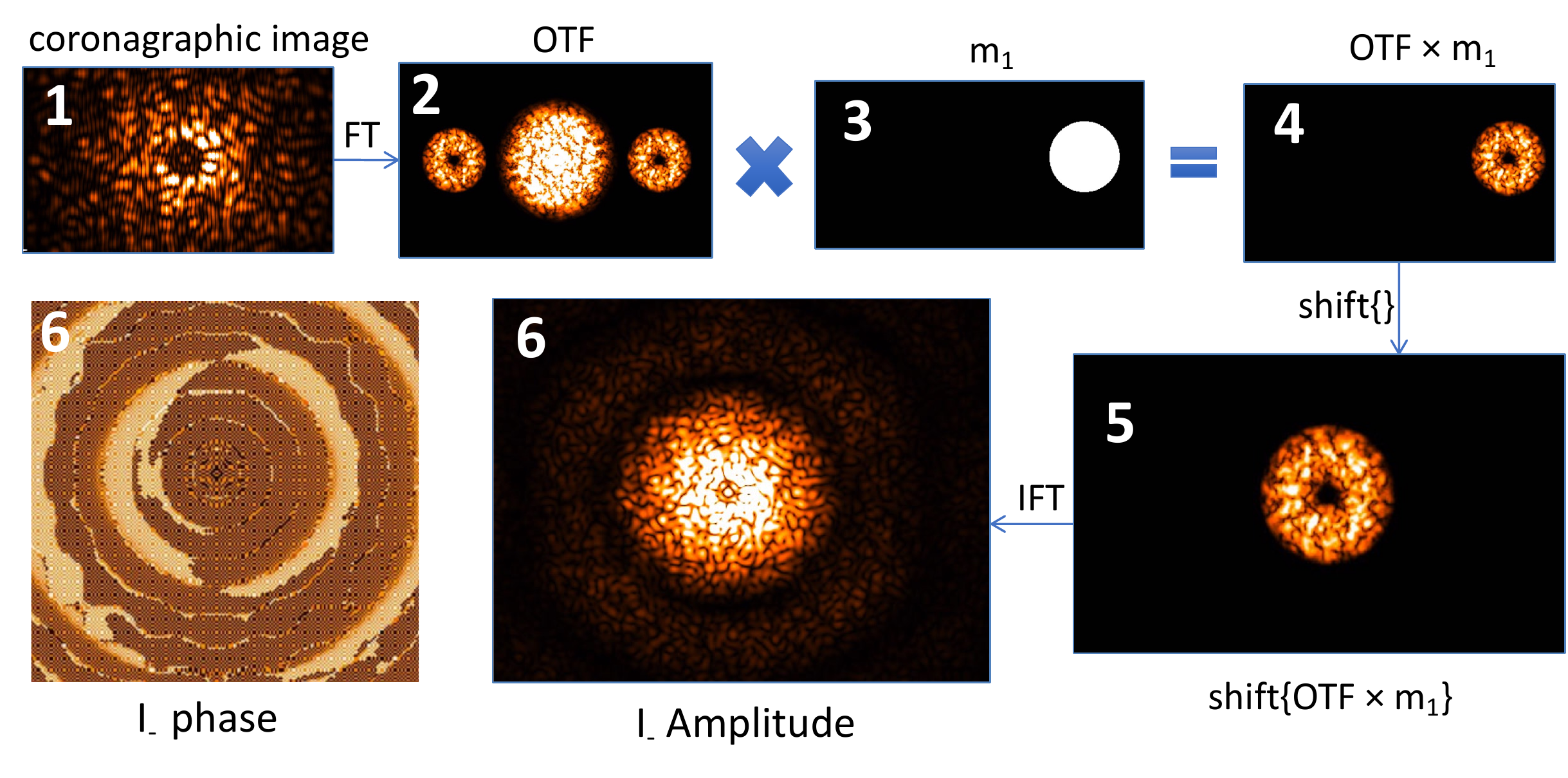}
    \caption[SCC Wavefront Sensing \& Reconstruction]{
    An illustration of the SCC wavefront sensing and reconstruction algorithm, adapted from Ref. \citenum{fast18}. Starting from the recorded coronagraphic SCC image (step 1), a series of Fourier Transforms (FT and IFT, the ``I'' representing ``inverse'') and Fourier filtering steps (steps 2-5) enable a complex-valued measurement of the focal plane electric field (step 6) known as $I_-$. Note that m$_1$ is a 2D array of ones and zeros (black and white representing zeros and ones, respectively; subsequently referred to as an ``algorithmic'' or ``binary'' mask), and the ``$\times$ symbol'' represents a multiplication of OTF (i.e., including both real and imaginary components) by m$_1$.
    }
    \label{fig: scc_wfsing}
\end{figure}

Fig. \ref{fig: scc} illustrates the concept of SCC image formation, both from a geometric (a) and Fourier (b) optics perspective.
Reviewing a mathematical description of the SCC image, the recorded image on the focal plane detector, $I$, produced from noiseless (i.e., without photon/detector/sky background noise) propagation of WFE (i.e., from atmospheric and/or instrumental origin) through to the detector, can be described by\cite{scc_orig},
\begin{equation}
    I=I_S\{\theta\}+I_P\{\theta\}+I_R\{\theta\}+2\sqrt{I_S\{\theta\} I_R\{\theta\}}\;\mathrm{cos}[2\pi(\xi_0/\lambda)\phi\{\theta\}],
    \label{eq: scc}
\end{equation}
where $\theta$ represents the two-dimensional coordinates in the image, $I_S$ is the un-fringed stellar speckle intensity component, $I_P$ is the un-fringed planet intensity component, $I_R$ is the pinhole point-spread-function (PSF) component (i.e., the PSF recorded if the main pupil aperture was blocked), and $\phi$ is the electric field phase difference between the $I_S$ and $I_R$ image components. The spatial scale of the fringe term $2\sqrt{I_S I_R}\mathrm{cos}[2\pi(\xi_0/\lambda)\phi]$ is set by the Lyot stop's pupil-to-pinhole separation ($\xi_0$, projected onto the entrance pupil) and observing wavelength, $\lambda$, enabling sub-$\lambda/D$ spatial scales (between $\lambda/[0.5 \xi_0]$ and $\lambda/[1.5 \xi_0]$) to be be isolated in the Fourier plane of the image (i.e., the optical transfer function, or OTF) without bias from the un-fringed exoplanet signal $I_P$. Fig. \ref{fig: scc_wfsing} illustrates this Fourier filtering process used to isolate the fringed SCC image component.
The phase of a given stellar speckle (i.e., $\phi\{\theta\}$ in Equation \ref{eq: scc}), which is not normally measureable in a single coronagraphic image (i.e., not encoded in $I_S$ or $I_R$), is encoded by the relative position of fringes on that speckle, which is measured as the phase component of $I_-$ in Fig. \ref{fig: scc_wfsing}. In equation form, the Fourier filtering process in Fig. \ref{fig: scc_wfsing} is
\begin{equation}
    \label{eq: I_minus}I_-=\mathrm{IFT}\{\mathrm{shift}\{\mathrm{FT}\{I\}m_1\}\} \\
\end{equation}
where Equation \ref{eq: I_minus} and Fig. \ref{fig: scc_wfsing} use the same operators for Fourier transform(FT\{\}), inverse Fourier transform (IFT\{\}), and OTF side-lobe centering (shift\{\}) as well as a binary mask to isolate the OTF side-lobe fringe component (m$_1$).
\subsection{DM Control}
\label{sec: scc_dm}
\begin{figure}[!h]
    \centering
    \includegraphics[width=0.7\textwidth]{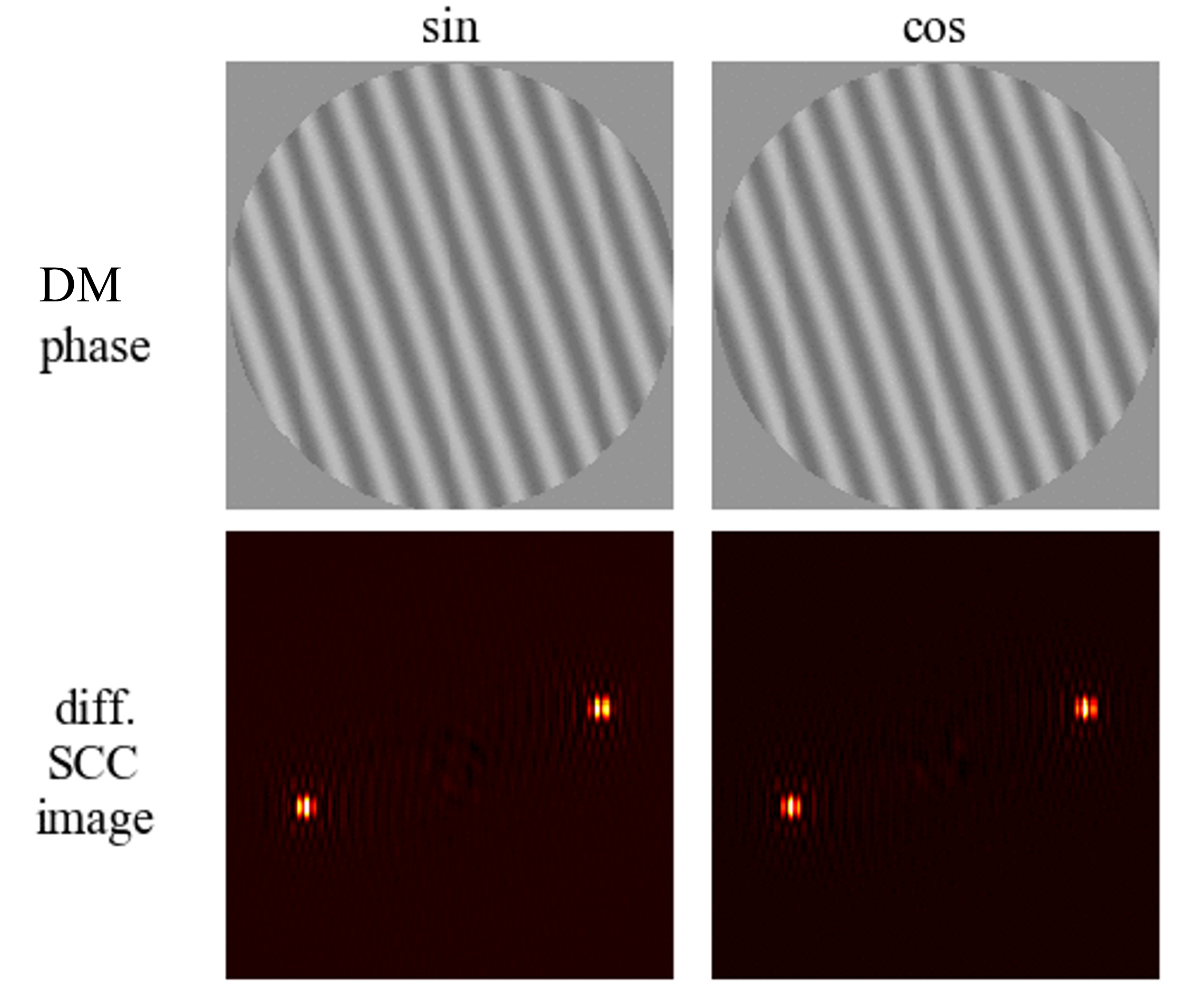}
    \caption[SCC interaction matrix generation]{
    An illustration of the procedure for generating a SCC interaction matrix (IM). The top row shows the phase applied on the DM, while the bottom row shows the corresponding differential SCC image (recorded image relative to a flat wavefront image). The two columns show responses for sine and cosine waves at the same spatial frequency. Such differential SCC images are recorded at the full range of controllable DM spatial frequencies to generate an IM in the complex-valued $I_-$ plane (see Fig. \ref{fig: scc_wfsing}).
    }
    \label{fig: scc_IM}
\end{figure}

Recording SCC images and processing them to produce $I_-$ can be calibrated empirically to deformable mirror commands applying sine and cosine waves over the full spatial frequency range of the DM, illustrated in Fig. \ref{fig: scc_IM}.
The Lyot stop pinhole size sets the image-plane region over which DM actuators can correct for speckles, so a higher order correction requires a smaller pinhole. A typical SCC pinhole diameter requirement is that the diagonal along the DM Nyquist region lie with the first minimum of the $I_R$ Airy function, which for a square grid DM is given in Ref. \citenum{mazoyer2013} (eq. 40).

\begin{figure}[!h]
    \centering
    \includegraphics[width=1.0\textwidth]{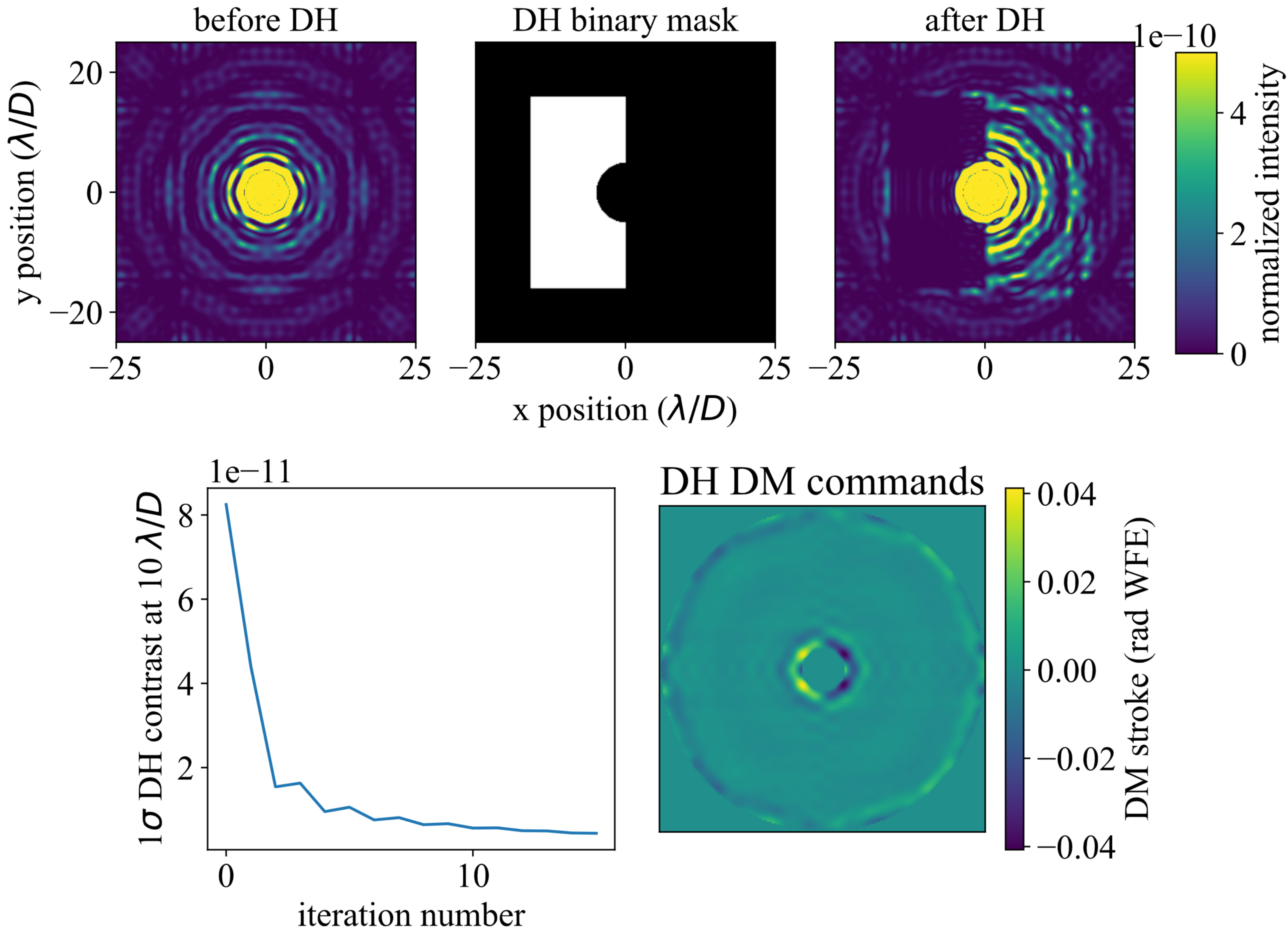}
    \caption[SCC DH illustration]{
    A simulated half DH on a diffraction-limited (i.e., no WFE) coronagraphic image (using the apodized Lyot coronagraph from Ref. \citenum{gpi_aplc}). Upper left: the raw APLC SCC image (the fringe visibility is too low to see fringes by eye). Upper middle: the binary mask (i.e., defining the DH region with ones and elsewhere with zeros) used to generate $m$ $I_-$ pixel values. Upper right: image after DH generation, after 15 iterations (the upper left and right panels are shown on the same scale). Lower left: contrast within the DH at 10 $\lambda/D$ vs. iteration number, showing that the majority of contrast gains occur after just 2 iterations. Lower right: DM commands used to generate the upper right DH image, illustrating the relative minimal amount of DM stroke needed for this SCC configuration.
    }
    \label{fig: aplc_dh}
\end{figure}

A SCC interaction matrix (IM) can be built by recording differential SCC images as shown in Fig. \ref{fig: scc_IM} for each DM Fourier mode (Eq. \ref{eq: scc} for a given DM Fourier minus Eq. \ref{eq: scc} for a flat DM) that are then converted to $I_-$ (Eq. \ref{eq: I_minus}) and lastly multiplied by an binary ``dark hole'' (DH) mask.\footnote{This step, not illustrated in Fig. \ref{fig: scc_IM}, is completed by multiplying (1) the complex-valued $I_-$ for a differential SCC image of a given DM Fourier mode by (2) an binary DH mask (i.e., using ones to represent the region of desired image-plane speckle attenuation). Note that with only a pupil plane conjugated DM(s), speckle correction of phase and amplitude errors (including diffraction) is limited to a half DH\cite{speckle_nulling}.} A vectorized version of these $I_-$ dark hole pixel values are then used to build an IM and then command matrix (CM) to enable a subsequent linear least-squares-based matrix vector multiplication (MVM) linking SCC images and DM commands as follows:
\begin{enumerate}
    \item 
    Applying $n$ Fourier modes (which includes sine and cosine components), recording $m$ $I_-$ pixel values (which includes the real and imaginary components) within the DH for each mode, a SCC reference matrix, $\vec{A}$, is generated with dimensions $n\times m$. A separate DM modal reference matrix, $\vec{F}$, with dimensions $n\times p$ (where $p$ is the number of DM actuators), is saved during this process.
    \item
    A CM, $\vec{C}$, is then generated via
    \begin{equation}
        \vec{C} = \vec{F}^T \cdot \left(\left((\vec{A}\cdot \vec{A}^T)^\dagger\right) \cdot \vec{A}\right),
        \label{eq: cm}
    \end{equation}
    where $\cdot$, $^T$, and $^\dagger$ represent a matrix dot product, transpose, and pseudo-inverse, respectively.
    %To enable modal gain optimization (i.e., a different loop gain is applied for different Fourier modes, but at the cost of two real-time MVMs insead of one), the CM is instead 
    %\begin{equation}
    %    \vec{C_2} = \left((\vec{A}\cdot \vec{A}^T)^\dagger\right) \cdot \vec{A}, 
    %\end{equation}
    \item 
    In closed-loop the $p\times m$ 
    %($\vec{C}$) or $n\times m$ ($\vec{C_2}$) 
    CM is then used to convert SCC $I_-$ DH pixel values for a new target image ($\vec{T}$, which is a $m\times 1$ vector) into DM commands ($\vec{D_n}$; a $p\times 1$ vector, where $\vec{D_{n-1}}$ represents the previous DM commands), via
    \begin{align}
        \label{eq: dmc} \vec{D_n}&=g(\vec{C} \cdot \vec{T})+l(\vec{D_{n-1}})%\mathrm{, or} \\
         %&= \left(\vec{g} \left(\vec{C_2} \cdot \vec{T}\right)\right)^T \cdot \vec{F} + l(\vec{D_{n-1}}), 
    \end{align}
    where $g$ is a scalar gain value (i.e., the same for all modes) 
    %, $\vec{g}$ is a $n\times 1$ modal gain vector (i.e., enabling different gains for each mode), 
    and $l$ is the scalar leak value for a leaky integrator closed-loop controller.
\end{enumerate}
This process to generate DM commands is then applied iteratively in closed-loop until desired DH contrast convergence in contrast achieved. The number of iterations needed to reach such is convergence system-dependent (e.g., coronagraph type, input WFE level, desired convergent DH level, available DM stroke, temporal WFE/stability environment, and more can all influence convergence timescale and achievable contrast) but can in principle be as few as 0 (i.e., only one image). Fig. \ref{fig: aplc_dh} illustrates this DH generation process.
Note that in the context of this section serving as a high-level summary of SCC DM control, many additional low-level details used for further improving achievable contrast are not described in this appendix section, such as optimizing/tuning the CM pseudo-inverse singular value decomposition (SVD) cutoff, IM Fourier modal amplitude, additional IM binary masks, the optimal basis set for low order modes, and more.
\section{
SEAL Alignment and Calibration
}
\label{sec: appendix_alignment_and_calibration}
Here we discuss SEAL-specific design and operations as it relates to the main text.
\subsection{FPM Alignment}
\label{sec: appendix_fpm_alignment}
FAST FPM alignment is carried out in multiple steps: initial coarse adjustment with a 5 degree-of-freedom stage (x-y-pitch-yaw-roll; focus is pre-determined by a knife edge test as described above). With the Lyot stop in place, pitch and yaw (i.e., tip and tilt) of the FPM mount are adjusted until the non-coronagraphic pupil aperture is centered on the intended central aperture of the Lyot stop (which displays a 0.8mm-wide ring around the aperture edge due to the 90\% undersizing). Then a pinhole aperture is placed on the focusing lens just upstream of the FPM to simulate a flat field. The FPM core in this mode is visible as a dark circular shadow on the otherwise illuminated flat field; x and y positions are then adjusted so that the center of this black dot is positioned at the coordinates of the normal Airy disk's center. 
After removing the pinhole/flatfield, a somewhat-aligned coronagraphic pupil is visible, by eye, on the Lyot stop (as in Figures \ref{fig: fast_sum} and \ref{fig: lab_setup}); further x-y adjustment can then enable visually maximizing the off-axis pupil intensity into the pinhole. The FPM roll is then adjusted to the desired position angle (i.e., defined by the Lyot stop pinhole's position angle), after which the above x-y positioning steps are repeated (since the setup-specific shape on the FPM is not exactly in the center of the optic's clear aperture). Lastly, fine x-y adjustment is performed in software with the DM tip/tilt: a grid search is performed in a square region around the estimated best alignment position, optimizing a combination of maximal fringe visibility \textit{and} minimal raw contrast.
\footnote{
Although a fringe visibility metric naively seems like a sufficient metric to align the FAST FPM (since once fully aligned the maximal amount of light is transmitted through the Lyot stop off-axis pinhole), fringe intensities are equally weighted between the focal plane amplitudes from the Lyot stop pinhole \textit{and} pupil (see Equation \ref{eq: scc}), which causes fringe intensities to increase when the FPM is misaligned because the amount of starlight increase is greater than the amount of pinhole light decrease. This is mitigated as described above by requiring minimal starlight (measured in the image plane) in addition to maximal fringe visibilities (measured by integrated flux on the MTF sidelobe).
} 
\subsection{Lyot Stop Design}
\label{sec: appendix_lyot_design}
The pupil-to-pinhole separation for our Lyot stop prescription is determined empirically, as was done in Ref. \citenum{fast_ne} and as we will illustrate should be a standard process for future FAST FPMs for laboratory and observatory instrument projects, by taking coronagraphic pupil images with the Lyot stop out and fitting the centers of each pupil (where a separate system is used with DMs replaced by fold mirrors for this characterization to enable a lower WFE characterization).\footnote{As to be discussed next in \S\ref{sec: appendix_calibration}, we are only able to reach 40 nm rms closed-loop WFE with the ALPAO DM, while in a separate system with only fold mirrors produces $\sim$nm rms level WFE, making the latter more optimal to characterize the quality of the fabricated FAST FPM (i.e., avoiding ambiguity issues between system WFE and FPM-specific fabrication defects.
} The off-axis pupil center is determined by fitting a two-dimensional Gaussian, while the central coronagraphic pupil position is determined by eye (incurring $\sim$1-2 pixel error $\approx$  2\% pupil diameter error $\approx$ 16 \% pinhole diameter error). Measuring this separation is important to enable a Lyot stop prescription that both (1) optimizes fringe visibility (i.e., limiting the pinhole position to deviate from the peak intensity), and (2) prevents a differential tilt between the pinhole PSF center and unfringed coronagraphic image center (i.e., preventing an offset between the center of the fringed and unfringed image components). We initially found that a Lyot stop prescription with a pinhole separation too discrepant from what was needed indeed caused a tilted pinhole PSF with respect to the unfringed coronagraphic image due to an amplitude gradient across the pinhole aperture causing a tilt. Although our initial pinhole separation prescription was for the intended FPM design, as discussed in \S\ref{sec: setup} it turned out that this differed significantly enough from the fabricated mask to cause such a measurable discrepancy, but with an empirically measured Lyot stop prescription this was no longer a problem. A recorded coronagraphic pupil image should therefore be the main determinant for future SCC Lyot stop prescriptions, both for lab experiments and future FAST instruments, to avoid a tilted pinhole PSF.
\subsection{Initial Calibrations}
\label{sec: appendix_calibration}
We start bench operations by flattening the ALPAO using the SHWFS, also compensating for any corresponding common path aberrations.
\footnote{
In reality the ALPAO surface is not flat, but instead compensates for the unflat resting MEMS and IrisAO surfaces. Although the latter produces an 8 nm rms flat as measured by our Zygo interferometer (and so likely further compensation by the ALPAO as measured by the SHWFS is negligible), we chose to keep the MEMS in a resting non-flat position while flattening the ALPAO, enabling ALPAO compensation for MEMS aberrations below 5.5 cycles/pupil (c/p). Although this option leaves the remaining MEMS figure errors between 5.5 and 12 c/p (the limiting spatial frequency on a side we correct for) initially uncorrected, we found this approach more favorable than using an optical flat command for the MEMS. We found that the latter approach---which we also generated with our Zygo interferometer---reaches diffraction-limited contrasts in the SCC image, requiring more MEMS stroke to dig a dark hole on diffracted light rather than on unpinned speckles, consistent with Ref. \citenum{pueyo09} (although note the differing contrast regimes between Ref. \citenum{pueyo09}, at $10^{-10}$ levels, and this paper, at $10^{-4}$ levels). Instead, leaving the MEMS figure error initially uncorrected at these mid to high spatial frequencies significantly relaxed the subsequent MEMS dark hole stroke limits (and also enabling the subsequent application of AO residual turbulence without running out of stroke) while having a negligible effect on initial Strehl ratio, since aberrations at these spatial frequencies are a negligible component of the overall wavefront error.
}
Similar ALPAO devices have been known to drift on hours to days timescales\cite{bitenc14}; we found that daily generation of a new best flat delivers sufficient performance. The flattening is done in SHWFS slope space, applying a classical zonal ``push minus pull'' actuator pokes to generate an IM\cite{hardy}. The CM pseudo inversion SVD cutoff is chosen to optimize the convergent closed-loop WFE as measured by the pre-calibrated SHWFS wavefront reconstructor (i.e., we choose the SVD cutoff value that produces the flattest wavefront), usually converging to a residual system WFE of around 40 nm rms.\footnote{
There are likely a few key limitations to reaching a deeper closed loop residual WFE: (1) the presence of a pinned (bad) MEMS actuator near the pupil center, which we do not account for in SHWFS reconstruction (e.g., algorithmically masking or interpolating over this region of the pupil), and (2) segment gaps caused by the IrisAO DM, which we also do not algorithmically account for in the SHWFS interaction matrix. Prior to the IrisAO and MEMS installation into SEAL, the SHWFS driving the ALPAO typically converged to $\sim$ 5 nm rms.} Note that this SHWFS flattening procedure is completed before running the DM-based FPM alignment procedure described in \S\ref{sec: fpm_alignment}.

To calibrate contrast, with the best flat on the MEMS applied as described above, we tilt the PSF off the FAST FPM core to record an off-axis PSF, adjusting the exposure time to prevent the star from saturating. Coronagraphic images in units of normalized intensity and/or contrast curves shown later in this manuscript are derived from dividing the dark-subtracted coronagraphic image in ADUs by the aforementioned maximum flux of the (also dark-subtracted) off-axis stellar PSF, scaling to account for the different exposure times in the two images. We can also calibrate flux ratio of the off-axis planet with this approach: turning off the starlight source but leaving on the planet source and scaling for exposure time gives a normalized planet flux ratio of 2.4$\times10^{-3}$.

To find the coronagraphic image center and empirically calibrate the plate scale we place two sine waves simultaneously on the DM--one with a position angle offset 90 degrees from the other and each at separations of 10 $\lambda/D$--to generate ``satellite spots.''\footnote{Viewing aligned coronagraphic pupil images and poking MEMS actuators confirmed the 26 actuator sampling across the Lyot stop pupil described in \S\ref{sec: setup}, which is needed to ensure a 10 c/p sine wave places spots at separations of 10 $\lambda/D$ in the SCC image. We also determined that the MEMS actuator grid has a 1$^\circ$ position angle offset with respect to the Andor detector pixel grid, likely due to small differences in how the two devices are mounted. We correct for this algorithmically to ensure that the MEMS Fourier modes are aligned with the Andor camera pixel grid.} Averaging the four 2D Gaussian-fit positions of these spots provides the coordinates of the image center. The average separation between the sine spot pairs also provides a plate scale measurement: 6.2 pixels/resel, sufficiently oversampled for the SCC fringes.

A calibration between DM voltage units and reconstructed (reflected) wavefront units for both the ALPAO and MEMS, which we use subsequently in this manuscript, were obtained by measuring 4 off-axis PSFs, each at four different combinations of tip and tilt commands (repeating this for both the ALPAO and MEMS). The above-calibrated plate scale is then used in combination with the fitted positions of the four off-axis PSFs to convert differential tip and tilt commands into separations in $\lambda/D$, which relates to pupil peak-to-valley (PV) tip/tilt WFE via: WFE/$\lambda=X(\lambda/D)$, where $X$ is the measured separation in the focal plane in $\lambda/D$ units. Averaging the separate tip and tilt differences produces 8 and 55 $\mu$m of PV WFE for 1 MEMS and ALPAO unit, respectively (i.e., the tip/tilt stroke limits of these devices).\footnote{It is important to note that in reality these conversion factors are mode-dependent. E.g., the ALPAO DM waffle stroke limit is  4 $\mu$ PV WFE, not 55 (ALPAO, private communication). However, because the main focus of this paper is the first laboratory demonstration of FAST correction of dynamic AO residual wavefront errors rather than high accuracy wavefront reconstruction, for simplicity we use these scalar conversion factors for all analyses and plots in this paper.}
%
% References
\bibliography{report} % bibliography data in report.bib
\bibliographystyle{spiejour}   % makes bibtex use spiejour.bst

\listoffigures
\listoftables
\begin{comment}
\end{comment}
\end{spacing}
\end{document}